\documentclass[]{jfm}
\usepackage{graphicx}
\usepackage{dcolumn}
\usepackage{bm}
\usepackage[utf8]{inputenc}
\usepackage{float}
\usepackage{soul}
\usepackage[utf8]{inputenc}
\usepackage[T1]{fontenc}
\usepackage{mathptmx}
\usepackage{etoolbox}
\usepackage{xcolor}
\usepackage{epstopdf,epsfig}
\usepackage[justification=justified,format=plain, width=\textwidth]{caption}
\usepackage{newtxtext}
\usepackage{newtxmath}
\usepackage{natbib}
\usepackage{hyperref}
\hypersetup{
    colorlinks = true,
    urlcolor   = blue,
    citecolor  = blue,
}

\title{Cross-variable amplitude-frequency coupling during intermittency in a turbulent thermoacoustic system} 

\author{Shruti Tandon \aff{1}, Aswin Balaji \aff{1}, Rohit Radhakrishnan \aff{1}, Manikandan Raghunathan \aff{1}, Gaurav Chopra \aff{1} and R. I. Sujith \aff{1}
  \corresp{\email{sujith@iitm.ac.in}}}

\affiliation{\aff{1}Department of Aerospace Engineering, Indian Institute of Technology Madras, Chennai 600 036, India\\
Centre of Excellence for Studying Critical Transitions in Complex Systems, Indian Institute of Technology Madras, Chennai 600 036, India}

\begin{document}
\maketitle

\begin{abstract}

We investigate flame-acoustic interactions in a turbulent combustor during the state of intermittency before the onset of thermoacoustic instability using complex networks. Experiments are performed in a turbulent bluff-body stabilized dump combustor where the inlet airflow rate is varied (a) quasi-statically and (b) continuously. We construct a natural visibility graph from the local heat release rate fluctuations ($\dot{q}’$) at each location. Comparing the average degree during epochs of high- and low-amplitude acoustic pressure oscillations ($p'$) during the state of intermittency, we detect frequency modulation in $\dot{q}’$. Through this approach, we discover unique spatial patterns of cross-variable coupling between the frequency of $\dot{q}’$ and the amplitude of $p'$. The frequency of $\dot{q}’$ increases in regions of flame anchoring owing to high-frequency excitation of the flow and flame during epochs of high-amplitude $p’$ dynamics. On the other hand, the frequency of $\dot{q}’$ decreases in regions associated with flame front distortions by large coherent vortices. In experiments with continuously varying airflow rates, the spatial pattern of frequency modulation varies with an increase in the average amplitude of $p’$ owing to an increase in the epochs of periodic $p’$ dynamics and the size of vortices forming in the flow. Dynamic shifts in the location of flame anchoring induce low-frequency fluctuations in $\dot{q}’$ during very high-amplitude intermittent $p’$ dynamics. Our approach using conditional natural visibility graphs thus reveals the spatial pattern of amplitude-frequency coupling between the co-evolving flame and the acoustic field dynamics in turbulent reacting flows.
\end{abstract}






\section{Introduction}

Complex systems such as turbulent thermo-fluid systems comprise a web of interactions within and between multiple subsystems, and across several spatial and temporal scales \citep{kauffman1995home, bertalanffy1968general, ottino2003complex, ladyman2013complex}. These interactions manifest as intriguing patterns, or as coupling of physical variables of the same or different subsystems \citep{estrada2023complex}. In order to predict or control the dynamics in such systems, it is essential to characterize the emergent pattern and cross-variable couplings. In this work, we investigate the spatial pattern of coupling between the acoustic and combustion dynamics in a turbulent thermoacoustic system that is ubiquitous in combustion engines in the aviation, space and power industries. A turbulent thermoacoustic system consists of a reacting flow where combustion is confined to a duct and the acoustic field in the confinement interacts with the turbulent flame and flow dynamics. Turbulent thermoacoustic systems exhibit both chaotic as well as emergent order in the spatio-temporal dynamics  \citep{sujith2021thermoacoustic}. Occurrence of order in such systems is catastrophic and occurs when a positive feedback is established between the flame, fluid and acoustic field dynamics \citep{sujith2021thermoacoustic, lieuwen2021unsteady}. During this state, referred to as thermoacoustic instability, the acoustic pressure oscillations exhibit ruinously high-amplitude self-sustained periodic oscillations \citep{lieuwen2002experimental}. Prediction of the occurrence of thermoacoustic instability and its amplitude and frequency, and developing strategies for prevention or control of thermoacoustic instability are highly desirable in the aviation and power generation industries \citep{lieuwen2021unsteady}.

Acoustic oscillations grow in amplitude if the heat released due to combustion adds energy constructively to the acoustic field. When the maximum rate of heat released due to combustion coincides with the maximum acoustic pressure fluctuations at several locations in the flow field, then, the net acoustic driving becomes greater than the acoustic damping and the system attains thermoacoustic instability \citep{rayleigh1878explanation, putnam1964general}. We note that, during this state, the underlying flow as well as flame dynamics are turbulent, and yet the acoustic pressure exhibits self-sustained limit cycle oscillations. Such emergence of order despite turbulent fluctuations has been viewed analogous to emergent periodicity during phase turbulence in a system of globally and diffusively coupled oscillators with cubic nonlinearity \citep{morales2024universality}. 

On the other hand, if the net damping is greater than acoustic driving, the combustor operates in a stable mode and exhibits low-amplitude acoustic fluctuations referred to as combustion noise \citep{strahle1978combustion}. Though referred to as noise, this state is characterized by high-dimensional deterministic chaos \citep{tony2015detecting, george2018pattern}, and is considered analogous to defect turbulence in a system of diffusely coupled nonlinear oscillators \citep{morales2024universality}. 
Transition from chaos (combustion noise) to order (thermoacoustic instability) in turbulent combustors occurs via the route of intermittency that is a state characterized by epochs of both periodic and chaotic spatio-temporal dynamics \citep{nair2014intermittency, george2018pattern}. 
During this state, bursts of periodic dynamics occur amidst epochs of aperiodic fluctuations repeatedly and in a self-sustained manner in the acoustic pressure signal. Clearly, there is a balance between the acoustic damping mechanisms and the acoustic driving through flame-acoustic interactions during the occurrence of intermittency. 

Then, an intriguing question is, how do the interactions manifest and self-organize? In this work, we investigate the flame-acoustic interactions during the state of intermittency using the framework of complex systems. In the following subsections, we discuss the current understanding of physical processes and interactions (Section \ref{sec_intro_interactions}), and the intermittency route of transition from chaos to order (Section \ref{sec_intro_intermittency}) in turbulent combustors. Next, we motivate the approach of complex systems theory for our purpose and describe the use of complex networks to study interactions in fluids and thermo-fluids (Section \ref{sec_intro_complexsys}).

\subsection{Physical processes and interactions in turbulent thermoacoustic systems}\label{sec_intro_interactions}

The structure, propagation and dynamics of flame in a turbulent combustor depends on the underlying turbulent flow, the type of flame (such as, premixed or partially premixed) and the flame stabilizing mechanism \citep{ballal1975structure, masri2015partial}. In turbulent combustors, the flame is stabilized by creating recirculation zones using a swirler, a bluff-body, a sudden expansion through a backward-facing step or a combination of these mechanisms. The flame is situated primarily along the mixing layers close to or in the recirculation zones. The flame is kept alive by auto-ignition kernels in flow pockets where the fuel-air mixture is propagated by the recirculating flow \citep{mastorakos2009ignition}. 

Combustion occurs in regions containing the most reactive mixture ratio which is determined by fine-scale mixing in the flow.
Large-scale distortions appear in a turbulent flame front created by vortices shed in the flow from the flame-stabilizer \citep{schadow1992combustion}. Vortices shed in the recirculation zone carry the fuel-air mixture downstream as they convect. Large-scale vortices delay fine-scale mixing and hence alter the location of intense combustion \citep{poinsot1987vortex, schadow1992combustion}. The flame anchoring and flame front fluctuations depend on the size and shedding frequency of vortices in the combustor, which in turn depend on the local acoustic velocity perturbations \citep{matveev2003model, seshadri2018predicting}. Thus, the flame structure is perturbed by the acoustic field. Moreover, the turbulent fluctuations in the flame determine the fluctuations in heat release rate that in turn influence the acoustic pressure fluctuations.

Flame and acoustic field dynamics are thus intrinsically coupled in a turbulent combustor \citep{chu1958non, lieuwen2021unsteady}. Extensive experiments and simulations have been performed in order to understand the coupling of periodic acoustic oscillations and turbulent flame dynamics during the state of thermoacoustic instability \citep{polifke2020modeling, noiray2008unified, schulz2018autoignition, silva2015numerical}. A common approach is to determine the flame transfer/ describing function (FTF/ FDF) that measures the heat release rate fluctuations in response to an external acoustic excitation upstream of the flame that is either harmonic or comprising a broadband of frequencies \citep{polifke2014black, schulz2018autoignition}. The FTF/ FDF is then used to determine the linear stability bounds of operating a combustor. 

\citet{shanbhogue2009flame} studied flame-sheet dynamics in response to external harmonic excitation in a turbulent bluff-body stabilized atmospheric burner and showed that the wrinkling of the flame surface increases with the amplitude of acoustic harmonic excitation. Similarly, \citet{balachandran2005experimental} performed experiments inducing acoustic excitation and perturbing the inlet flow in a turbulent premixed flame combustor. They showed that the amplitude as well as frequency of the acoustic excitation affected the vortex shedding frequency, shear layer roll up and flame response. Also, novel system identification approaches \citep{polifke2020modeling, gopinathan2018nonlinear, noiray2016Denisov} have been developed to identify the acoustic growth rates using reduced order models for flame frequency response, for example using the Fokker Plank formalism for describing the acoustic dynamics in thermoacoustic systems \citep{noiray2016Denisov}. Recently, \citet{passarelli2021cross} showed the presence of nonlinear coupling between the frequencies of oscillations in different variables such as acoustic pressure, heat release rate and velocity fluctuations in a model gas-turbine combustor.

\subsection{The intermittency route of transition from chaos to order}\label{sec_intro_intermittency}

In turbulent thermoacoustic systems, the intermittency route of transition from chaos to order exhibits loss of multifractality \citep{nair2014multifractality} in the temporal dynamics of acoustic pressure through a scaling law between the Hurst exponent and the normalized acoustic pressure amplitude \citep{pavithran2020universality}. Intriguingly, this scaling law is found to be universal across turbulent aero-acoustic, aeroelastic and thermo-fluid systems \citep{pavithran2020universality}. During the state of intermittency, the spatial field delineates a chimera state, i.e., the acoustic pressure and the heat release rate dynamics are intermittently synchronized in certain pockets of the spatial field, while at other locations these oscillations are desynchronized \citep{mondal2017onset}. Also, the flame exhibits a fractal organization where the wrinkles along the flame surface appear similar at multiple length scales \citep{raghunathan2020multifractal}. 

The interdependence between the acoustic pressure and the heat release rate fluctuations during the state of intermittency plays a crucial role in determining the qualitative features as well as the quantitative scaling relations of the dynamical transition. \citet{unni2017flame} showed that the flame in a turbulent combustor exhibits two types of oscillatory patterns during this state, i.e., aperiodic owing to the underlying turbulence and acoustic perturbations and periodic roll-up owing to the large-scale vortices shed in the dump plane.

Using synchronization theory, \citet{pawar2017thermoacoustic} showed that the acoustic pressure ($p'$) and the global heat release rate ($\dot{q}'_G$) transition from a de-synchronized state to an intermittently phase synchronized state to a generalized synchronized state as the system dynamics transition from chaos to order via the route of intermittency. \citet{godavarthi2018coupled} established that a strong interdependence exists between  $p'$ and $\dot{q}'_G$ during the states of intermittency and order (thermoacoustic instability) in the system using cross and joint recurrence analysis. \textcolor{black}{\citet{tandon2020bursting} presented a phenomenological model with coupling between slow and fast timescales corresponding to distinct subsystems to replicate amplitude-modulated bursting observed during the state of intermittency in practical low-turbulence combustion systems.} In the model, the authors proposed that the amplitude of acoustic pressure and frequency of heat release rate fluctuations are coupled. Furthermore, the typical pattern of interactions between the acoustic, combustion and flow subsystems is distinct \citep{tandon2023multilayer} during chaotic, intermittent and ordered dynamical states in turbulent thermoacoustic systems. \citet{tandon2023multilayer} showed that localised pockets of interactions between the fluid and acoustic-coupled heat released rate dynamics emerge during the state of intermittency and foster self-organized feedback between subsystems leading to the emergence of self-sustained order.


\subsection{Adopting complex systems approach}\label{sec_intro_complexsys}
The coupling of the physical processes, such as that between flame and acoustic, is traditionally viewed as a two-way causality. That is, variations in the flame dynamics affect the acoustic field and vice versa. However, the interdependence between the two processes can be viewed as a \textit{holistic causation} of the system of flame-flow-acoustic interactions. The term \textit{holistic causation} implies that the two subsystems are essentially co-evolving as a single entity and the cause and effect of each on the other cannot be separated \citep{bertalanffy1968general}. This view is inspired from complex systems theory \citep{ottino2003complex, siegenfeld2020introduction, estrada2023complex}, according to which the spatial structure of interactions, the function (physical processes associated with each subsystem) and the dynamics of the system are interdependent and non-separable \citep{witherington2011taking}. The nature and pattern of interactions determine the dynamics as much as the dynamics determines the pattern of interactions. Such interplay between the dynamics and patterns is called higher-order interaction and a topological explanation of features and patterns is evoked which facilitates an understanding of the processes ensuing in the system \citep{estrada2023complex, arshinov2003causality}.

A complex system \textcolor{black}{can be} represented and analysed in the form of a complex network. A complex network consists of nodes connected by links, where links are used to represent interactions between different subsystems or spatial locations (represented by nodes). Networks can be constructed from time series using different methods such that specific features of the time series are mapped onto the topology of the network \citep{gao2017complex}. The framework of complex networks has been robust in capturing a variety of features observed during the dynamical transitions in fluid and thermo-fluid systems \citep{sujith2020complex, iacobello2021review, iacobello2021large, taira2022network}. \textcolor{black}{Networks constructed from the time series of a system variable have been used extensively to study dynamical transitions in turbulent thermoacoustic systems \citep{murugesan2015combustion, godavarthi2017recurrence, aoki2020dynamic, gotoda2017characterization, okuno2015dynamics, tandon2021condensation}, noise-perturbed laminar jets \citep{guan2023multifractality}, and self-excited jets \citep{murugesan2019recurrenceinjet}. Network analysis of time series has also helped identify synchronization between the acoustic pressure and the heat release rate fluctuations during different dynamical states in thermoacoustic systems \citep{godavarthi2018coupled, kobayashi2019early}.}

Networks constructed from the spatial field of fluid flows have provided both insights into the physical processes ensuing in the system and the means to identify sensitive locations suitable for implementing smart control strategies \citep{nair2018networkcontrol, yeh2021network, meena2021identifying}. A time-varying network analysis of the spatial field of thermoacoustic power generation \citep{abin2019jfm} revealed that spatial pockets of coherent thermoacoustic power grow into large clusters as order emerges in the system. Novel methods of constructing spatial networks from vortical interactions in turbulent flows have been proposed recently \citep{taira2016network}, where a node is connected to another by the strength of velocity induced at the latter location by the vorticity of the former location and vice versa. In turbulent combustors, this approach was used to identify localized pockets of vortical interactions that can be perturbed via microjet air injections leading to efficient suppression of thermoacoustic instability \citep{hashimoto2019spatiotemporal, abin2021jfmtaira}. Moreover, by identifying communities of vortical interactions, \citet{sahay2023community} show that interactions in the flow are strongly correlated to acoustic pressure dynamics during the state of thermoacoustic instability. Further, higher-order interactions such as those between vorticity and acoustic-coupled combustion dynamics have been recently deciphered using multilayer networks \citep{tandon2023multilayer}. The coupling mechanism of acoustic and flame dynamics adds further complexity to the process of self-organized interactions within the system. 

\subsection{Questions addressed in this work:}

\textcolor{black}{A variety of intriguing features emerge in the spatio-temporal dynamics due to inter-subsystem interactions \citep{sujith2020complex, sujith2021thermoacoustic} and manifest as coupling of physical variables \citep{estrada2023complex}. Some important questions that remain unanswered are: How does the coupling between the physical variables of flame and acoustic subsystems manifest during the state of intermittency? How do these interactions vary spatially? How do the flame-acoustic interactions depend on the flow conditions such as variations in the inlet flow rates?} Understanding the spatial pattern of flame-acoustic coupling during intermittency prior to the onset of thermoacoustic instability is crucial during combustor design to control flame-acoustic interactions and hence increase operational margins of the combustor. 

In this work, we present a data-driven approach using complex networks to study the spatial structure of interactions between the acoustic pressure and the heat release rate dynamics. We perform experiments in a turbulent bluff-body stabilized combustor for two distinct inlet flow conditions, one where the inlet airflow rate is varied in a quasi-static manner, and another where the airflow rate is continuously changed with a finite rate (described in Section \ref{sec_expt}). We construct natural visibility graphs from the heat release rate fluctuations at each location in the combustor and study the variation in the network properties with respect to variations in the acoustic pressure dynamics (Section \ref{sec_dataAnalysis}). We show that the inter-subsystem interactions in a turbulent thermoacoustic system manifest as a unique pattern of cross-variable amplitude-frequency coupling between the acoustic field and the heat release rate fluctuations during the state of intermittency.


\section{Experiments in a turbulent thermoacoustic system}\label{sec_expt}

\begin{figure}
    \centering
    \includegraphics[width=\linewidth]{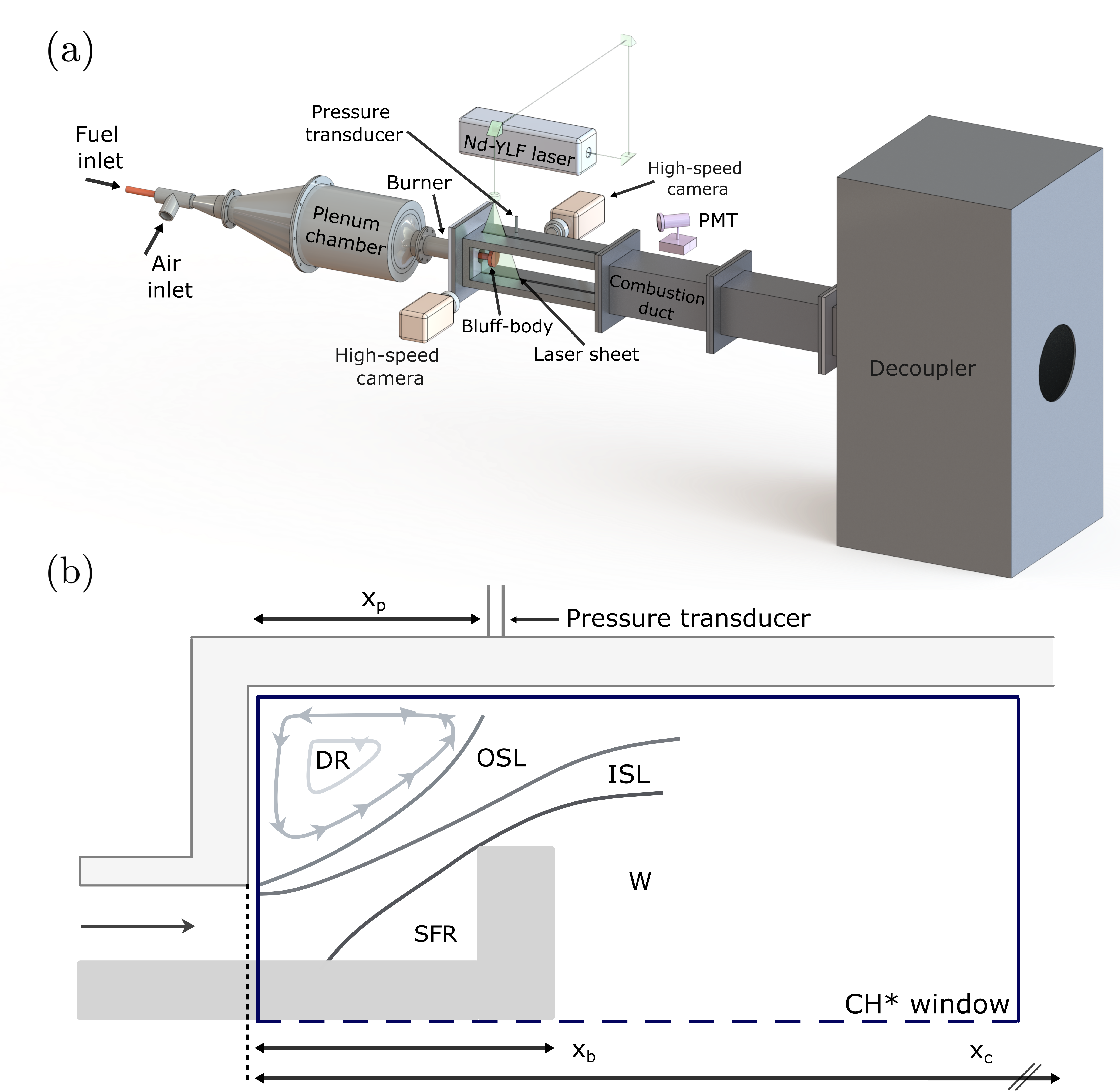}
    \caption{Schematic of (a) the experimental setup of a bluff-body stabilized turbulent combustor and (b) a cross-section of the combustor showing the backward-facing step, the bluff-body (gray object), the analysis window ($\text{CH}^*$ window). The different flow regions are labeled as follow: DR denotes the recirculation zone created in the dump plane downstream of the backward-facing step in the dump plane, OSL and ISL denote the outer and inner shear layer, SFR denotes the region of secondary flow recirculation created upstream of the bluff-body and W denotes the wake downstream of the bluff-body. The location of the pressure transducer from the backward facing step ($x_p$), length of the combustion chamber ($x_c$), and the bluff-body position from the dump plane ($x_b$) are marked in the schematic (b).}
    \label{fig_exptschematic}
\end{figure}

Experiments were performed in a laboratory-scale turbulent bluff-body stabilized dump combustor, a schematic of which is shown in figure \ref{fig_exptschematic}(a). Two set of experiments were conducted: (i) where a control parameter is varied in a quasi-static manner and (ii) involving continuous variation in a control parameter. The combustion chamber includes a sudden flow expansion using a backward-facing step. The flame is stabilized downstream of the backward-facing step using a circular bluff-body, which is 47 mm in diameter and 10 mm in thickness and is mounted on a 16 mm-diameter shaft. The combustion chamber has a cross-section of 90 $\times$ 90 $\text{mm}^2$. A decoupler measuring 1000 $\times$ 500 $\times$ 500 mm is attached to the end of combustion chamber. The decoupler ensures the pressure at the end of the duct is the same as the atmospheric pressure such that fluctuations at the boundary are negligible. The other details of the setup configuration such as the bluff-body location from the dump plane, combustor length, fuel injection location upstream of the dump plane from the backward facing step ($x_f$), and the decay rate of the setup are tabulated in table~\ref{table:exp1} for both sets of experiments. The acoustic damping was maintained within limits to ensure repeatability of the experiments (refer table \ref{table:exp1}). 

We use liquefied petroleum gas (LPG) (butane 60\% and propane 40\%) as the fuel for combustion and air as the oxidizer. Fuel is injected through the shaft and ejected via four holes of 1 mm diameter. The fuel-air mixture is partially premixed in the burner before being injected into the combustion chamber. The mass flow rates of the inlet fuel and air, measured in SLPM (standard litres per minute), are controlled using mass flow controllers (MFC). During experiments, we maintain the fuel flow rate constant and vary the air flow rate (flow control parameter), thus varying the Reynolds number ($Re$) of the inlet fuel-air mixture. The equation used to find $Re$ close to the burner position is given by $Re = 4\dot{m}D_1/\pi \mu D_0^2$, where $\dot{m}$ represents the total of the fuel flow rate ($\dot{m_f}$) and air-flow rate ($\dot{m_a}$), $D_0$ denotes the hydraulic diameter of the burner, $D_1$ is the circular bluff-body diameter, and $\mu$ is the kinematic viscosity of the fuel-air mixture under experimental conditions \citep{nair2014intermittency}. The $Re$ of a binary gas mixture was calculated using the dynamic viscosity expression given by \citet{wilke1950viscosity}. 

The acoustic pressure fluctuations ($p'$) in the duct is measured using a piezoelectric pressure transducer. The piezoelectric transducer is mounted on a pressure port (T-joint) that is flush mounted on the combustor wall. To protect the transducer from excess heating from the combustor, a Teflon adapter is used. Further, one shoulder of the T-joint was also provided with a semi-infinite tube 10 m in length to prevent acoustic resonance within the ports, thereby minimizing the frequency response of the system. The global unsteady heat release rate ($\dot{q}_G$) is measured using a photomultiplier tube (PMT) module. Also, we capture high-speed chemiluminescence images, which are then used to determine the local unsteady heat release rate ($\dot{q}(x,y)$) of the flame. The heat release rate fluctuations are calculated at each location as $\dot{q}'(x,y)=\dot{q}(x,y) - \langle\dot{q}(x,y)\rangle$, where $\langle \rangle$ denotes a time average. The specification of the instruments used for data acquisition and the operational parameters for both the experimental configurations are tabulated table~\ref{table:exp2}. \textcolor{black}{During experiments with continuous variations in the airflow rate, the rate of change of the equivalence ratio is $d\phi/dt=-0.0247$. This rate of change of equivalence ratio is close to that typically reported for gas-turbine engines \citep{davies1971handling, heywood1973parameters}. Recently,\cite{manikandan2020rate} have discussed the need to study rate-dependent bifurcations in order to determine the thermoacoustic stability map in a turbulent afterburner rig.}


\begin{table}
\centering
\def~{\hphantom{0}}
\begin{tabular}{>{\centering\arraybackslash}m{2cm}>{\centering\arraybackslash}m{4cm}>{\centering\arraybackslash}m{4cm}}
\multicolumn{1}{c}{\begin{tabular}[c]{@{}c@{}}System Specifications\end{tabular}} & \begin{tabular}[c]{@{}c@{}}Quasi-static experiments\end{tabular} & \begin{tabular}[c]{@{}c@{}}Continuous variation of\\ parameter experiments\end{tabular} \\ \hline
$x_b$                                                                                & 45 mm                                                               & 30 mm                                                                                    \\
$x_c$                                                                                & 1100 mm                                                             & 1400 mm                                                                                  \\
$x_f$                                                                                & 110 mm                                                              & 95 mm                                                                                    \\
Decay rate                                                                           & $19 \pm 1.0~\text{s}^{-1}$                                                           & $6.5 \pm 0.5~\text{s}^{-1}$                                            \\ \hline                 
\end{tabular}
\caption{System specifications for the turbulent bluff-body stabilized combustor used for experiments where the inlet airflow rate is varied either quasi-statically or continuously. Here, $x_b$ and $x_c$ represent the dimensions of the bluff-body location, and combustor length, respectively, as depicted in figure \ref{fig_exptschematic}. $x_f$ represents the location of fuel injection location upstream of the dump plane from the backward facing step.}
\label{table:exp1}
\end{table}

 

\begin{table}
\centering
\renewcommand{\arraystretch}{1.5}
\begin{tabular}{lcc}
\multicolumn{1}{c}{\begin{tabular}[c]{@{}c@{}}Equipment\\ details\end{tabular}}    & \begin{tabular}[c]{@{}c@{}}Quasi-static \\ experiments\end{tabular}                                              & \begin{tabular}[c]{@{}c@{}}Continuous variation \\ of parameter experiments\end{tabular}                \\ \hline
\multicolumn{3}{l}{\textbf{Piezoelectric pressure sensor}}                                                                                                                                                                                                                                                      \\
Make, uncertainty                                                                  & \multicolumn{2}{c}{PCB103B02, $\pm 0.15$ Pa}                                                                                                                                                                               \\
$x_p$                                                                              & \multicolumn{2}{c}{20 mm}                                                                                                                                                                                                  \\
\begin{tabular}[c]{@{}l@{}}Sampling frequency,\\ total operation time\end{tabular} & \begin{tabular}[c]{@{}c@{}}10 kHz\\ 3 s (for each equivalence ratio)\end{tabular}                                & \begin{tabular}[c]{@{}c@{}}10 kHz\\ 34 s (for total operation)\end{tabular}                             \\ \hline
\multicolumn{3}{l}{\textbf{Photomultiplier tube}}                                                                                                                                                                                                                                                               \\
Make                                                                               & \multicolumn{2}{c}{Hamamatsu H11462-012}                                                                                                                                                                                   \\
Filter                                                                             & \begin{tabular}[c]{@{}c@{}}$\text{OH}^{*}$ filter: 308 nm \\ with 12 FWHM\end{tabular}                           & \begin{tabular}[c]{@{}c@{}}$\text{CH}^{*}$ filter: 435 nm\\ with 12 FWHM\end{tabular}                   \\
Field of view                                                                      & \multicolumn{2}{c}{70$^\circ$}                                                                                                                                                                                             \\
\begin{tabular}[c]{@{}l@{}}Sampling frequency,\\ total operation time\end{tabular} & \begin{tabular}[c]{@{}c@{}}10 kHz\\ 3 s (for each equivalence ratio)\end{tabular}                                & \begin{tabular}[c]{@{}c@{}}10 kHz\\ 34 s (for total operation)\end{tabular}                             \\ \hline
\multicolumn{3}{l}{\textbf{Mass flow controllers}}                                                                                                                                                                                                                                                              \\
Make, uncertainty                                                                  & \multicolumn{2}{c}{Alicat Scientific MCR Series, $\pm$ (0.80 \% of the reading + 0.20 \% of the full scale)}                                                                                                               \\
\begin{tabular}[c]{@{}l@{}}Thermal power, \\ fuel mass flow rate\end{tabular}      & 43.6 kW, $(0.875 \times 10^{-3})$ kg/s                                                                           & 52.32 kW, $(1.050 \times 10^{-3})$ kg/s                                                                 \\
\begin{tabular}[c]{@{}l@{}}$Re$ range of air, \\ uncertainty\end{tabular}          & $1.25 \times 10^{4}$ to $1.8 \times 10^4$, $\pm 6\%$                                                             & $2.2 \times 10^{4}$ to $7.0 \times 10^{4}$, $\pm 6\%$                                                   \\
\begin{tabular}[c]{@{}l@{}}$\phi$ range, \\ uncertainty\end{tabular}               & 1.90 to 1.19, $\pm 0.02$                                                                                         & 2.162 to 0.552 ($d\phi / dt = -0.0247$), $\pm 0.02$                                                     \\ \hline
\multicolumn{3}{l}{\textbf{High-speed chemiluminescence images}}                                                                                                                                                                                                                                                \\
Camera                                                                             & \begin{tabular}[c]{@{}c@{}}High-speed CMOS \\ camera (Phantom – v 12.1)\end{tabular}                             & \begin{tabular}[c]{@{}c@{}}High-speed \\ CMOS camera (Phantom – VEO 710L)\end{tabular}                  \\
Filter                                                                             & \multicolumn{2}{c}{$\text{CH}^{*}$ filter: 435 nm,  90\% transmissivity}                                                                                                                                                   \\
Camera lens                                                                        & \begin{tabular}[c]{@{}c@{}}ZIESS 50 mm \\ lens at f/2 aperture\end{tabular}                                      & \begin{tabular}[c]{@{}c@{}}ZIESS 100 mm lens \\ at f/2 aperture\end{tabular}                            \\
\begin{tabular}[c]{@{}l@{}}Frame rate, resolution, \\ operation time\end{tabular}  & \begin{tabular}[c]{@{}c@{}}2000 fps, $800 \times 600$ pixels, \\ 2.5 s (for each control parameter)\end{tabular} & \begin{tabular}[c]{@{}c@{}}1250 fps, 640 $\times$ 480 pixels, 34 s\\ (for total operation)\end{tabular}
\\ \hline
\end{tabular}
\label{table:exp2}
\caption{Description and specifications of the instruments used for data acquisition during experiments with quasi-static and continuous variation of inlet airflow rate in a turbulent bluff-body stabilized combustor.}
\label{table:exp2}
\end{table}


Additionally, we performed Mie-scattering experiments for the same combustor configuration given in the quasi-static experiment. Mie scattering experiments were performed with olive oil droplets as seeding particles to qualitatively describe the flame front (\cite{zhang2010, SHANBHOGUE20091787}). In the current study, the mean diameter of the seeding particles used is approximately 1 $\mu$m at an equivalence ratio of 1.6. A Laskin nozzle is used to seed the olive oil droplets into the incoming air. These droplets scatter light when they pass through a laser sheet, but as the oil droplets traverse through the flame they evaporate and burn. The boundary of the illuminated area is marked as the flame front, since oil droplets are found exclusively in areas with unburned cold reactants. A Q-switched Nd:YLF laser (527 nm, in single pulse mode) synchronized with a high-speed camera (Phantom -- v 12.1) is used for the Mie scattering experiments. The camera is mounted with a ZEISS 100 mm lens with aperture at f/2 and outfitted with a short bandpass optical filter (527 $\pm$ 12 nm) to capture the scattered light from the seeding particles. The camera images $51.2~\text{mm} \times 36.8~\text{mm}$ between the dump plane and the bluff-body onto $1024 \times 736$ pixels of the camera sensor. The camera acquires the images at a frame rate of 5000 fps for a duration of 1.8 s.


Figure \ref{fig_exptschematic}(b) shows the window of analysis and a schematic of flow features in the field of the turbulent bluff-body stabilized combustor. As the flow is forced to turn around the backward-facing step and subsequently around the bluff-body, two recirculation zones are formed, one in the dump plane (region DR, figure  \ref{fig_exptschematic}(b)) and one in the wake of the bluff-body (region W, figure  \ref{fig_exptschematic}(b)), respectively. A secondary flow recirculation denoted as region SFR in \ref{fig_exptschematic}(b) is created upstream of the bluff-body where the flow along the shaft is forced to turn at a right angle by the bluff-body. Regions denoted as OSL and ISL in  \ref{fig_exptschematic}(b) represent the outer and inner shear layers respectively. The different flow regions will herewith be referred to in the manuscript by these abbreviations denoted in figure \ref{fig_exptschematic}(b). Also, we note that the acoustic pressure signal measured by the transducer at one location represents the acoustic pressure dynamics throughout the analysis window. This is because, an acoustic pressure anti-node (maximum of the acoustic mode in space) occurs around the location of the pressure transducer, and we find that the variations in $p'$ along the axial direction within our window of analysis are insignificant. \citet{nair2014role} and \citet{sudarsanan2023emergence} have confirmed this experimentally.

\subsection{Typical time series data during the state of intermittency}

\begin{figure}
    \centering
    \includegraphics[width=1\linewidth]{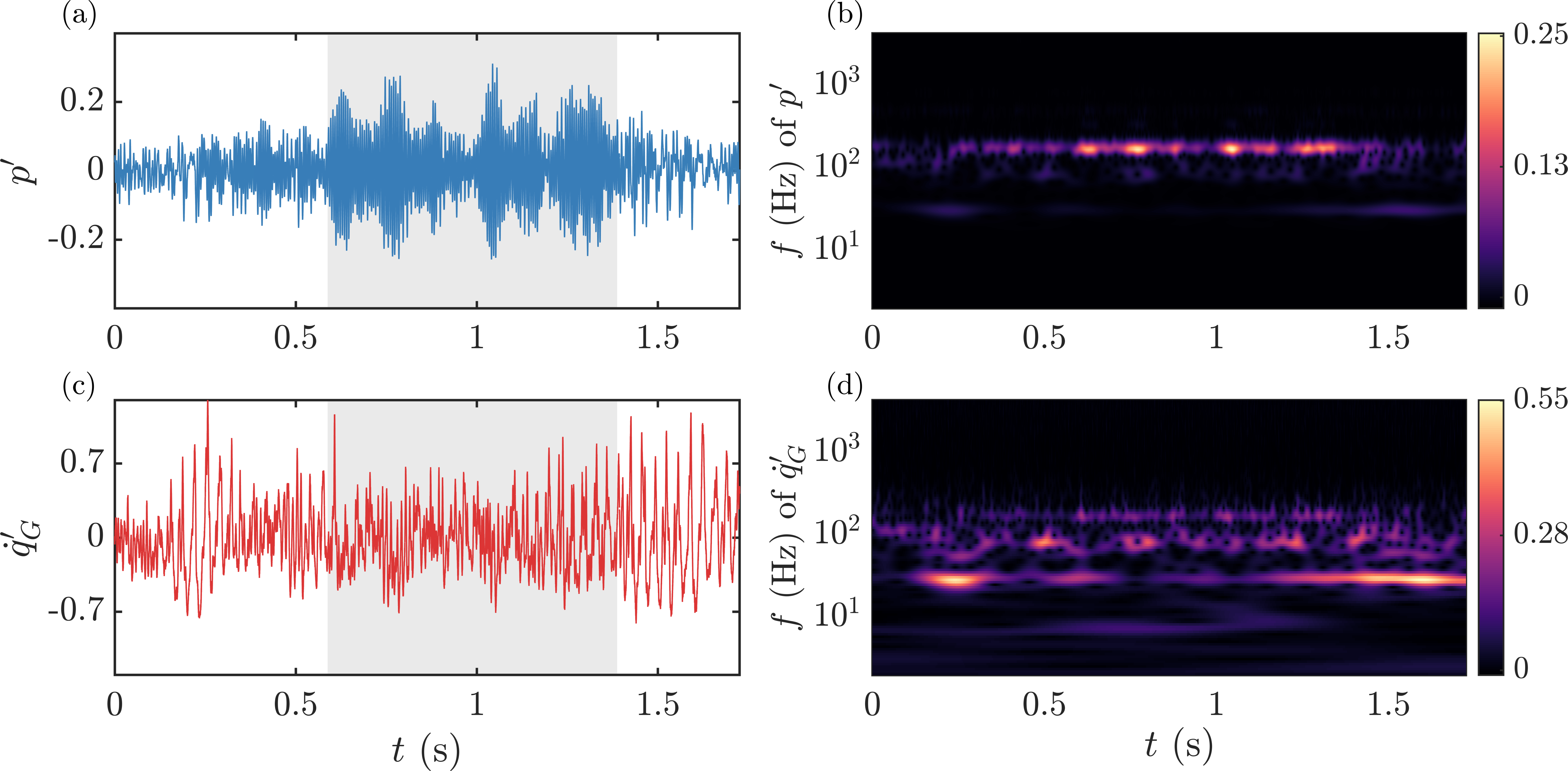}
    \caption{The time series (a, c) and the corresponding spectrograms showing the time variation of dominant frequencies  (b, d) in the acoustic pressure ($p'$) and global heat release rate ($\dot{q}_G'$) fluctuations, respectively during the state of intermittency. The shaded regions in the time series (a, c) correspond to the epochs of high-amplitude $p'$ dynamics. Comparing the time series and the spectrograms, we find low-frequency fluctuations in $\dot{q}_G'$ are predominant during low-amplitude epochs of $p'$ oscillations during the occurrence of intermittency in quasi-static experiments.}
    \label{fig_spectrogram}
\end{figure}

From quasi-static experiments, we observe the occurrence of intermittency en-route the transition from chaotic to periodic acoustic pressure dynamics in our system. Figure \ref{fig_spectrogram}(a,c) shows the time series of acoustic pressure ($p'$) and global heat release rate fluctuations ($\dot{q}_G'$), respectively, during the state of intermittency. Bursts of high-amplitude $p'$ oscillations are sustained for short epochs amidst aperiodic fluctuations in the acoustic pressure signal (figure \ref{fig_spectrogram}(a)). The $p'$ signal exhibits a dominant frequency around 180 Hz, as evident in the spectrogram in figure \ref{fig_spectrogram}(b). Also, the $\dot{q}_G'$ signal exhibits a dominant frequency at around 30 Hz during the epochs of low-amplitude $p'$-dynamics, while a band of dominant frequencies appears around 90 Hz during the epochs of high-amplitude $p'$ dynamics (see spectrogram in figure \ref{fig_spectrogram}(d)). 

Comparing the global time series and spectrograms obtained during the state of intermittency, we can infer that the frequency of the global heat release rate fluctuations ($\dot{q}_G'$) changes with variations in the amplitude of $p'$. However, significant amplitude modulations in $p'$ are not present during the state of combustion noise or thermoacoustic instability. In this work, we quantify the inter-dependence between the amplitude of acoustic pressure fluctuations and the frequency of heat release rate fluctuations and study the spatial pattern of such interactions during the state of intermittency. To do so, we use natural visibility graphs (NVG) constructed from the heat release rate fluctuations at each location and compare the network properties during high- and low-amplitude $p'$ dynamics.

\section{Data analysis \label{sec_dataAnalysis}}

\subsection{Constructing natural visibility graphs (NVG)}\label{sec_NVG_construct}

Natural visibility graph \citep{lacasa2008time,zou2019complex,nunez2012visibility} is a tool to encode the features of a time series onto a complex network comprising nodes and links. Each timestamp is considered as a node. A connection is established between two nodes if a straight line drawn between the corresponding data points does not intersect any other intermediate data point, i.e., the two nodes are `visible' to each other. Mathematically, node $i$ and $j$ of time series $y(t)$ will be connected to each other if every intermediate node $n$ satisfies the condition:

\begin{equation}
y(t_n) < y(t_j) + (y(t_i) - y(t_j)) \frac{t_j - t_n}{t_j - t_i}
\end{equation}

\vspace{5mm}

The network is represented by an adjacency matrix $\textbf{A} = [a_{ij}]$ of size $N \times N$, where $N$ is equal to the total number of time stamps in the time series. Elements of $\mathbf{A}$ are, $a_{ij} = 1$, if there exists a link between nodes $i$ and $j$, else $a_{ij} = 0$. Therefore, the adjacency matrix $\textbf{A}$ for NVG is symmetric, and the network is unweighted and undirected. Note that, the network generated is invariant under the scaling transformation of the time series \citep{lacasa2008time}. 

\begin{figure}
    \centering
    \includegraphics[width=1\linewidth]{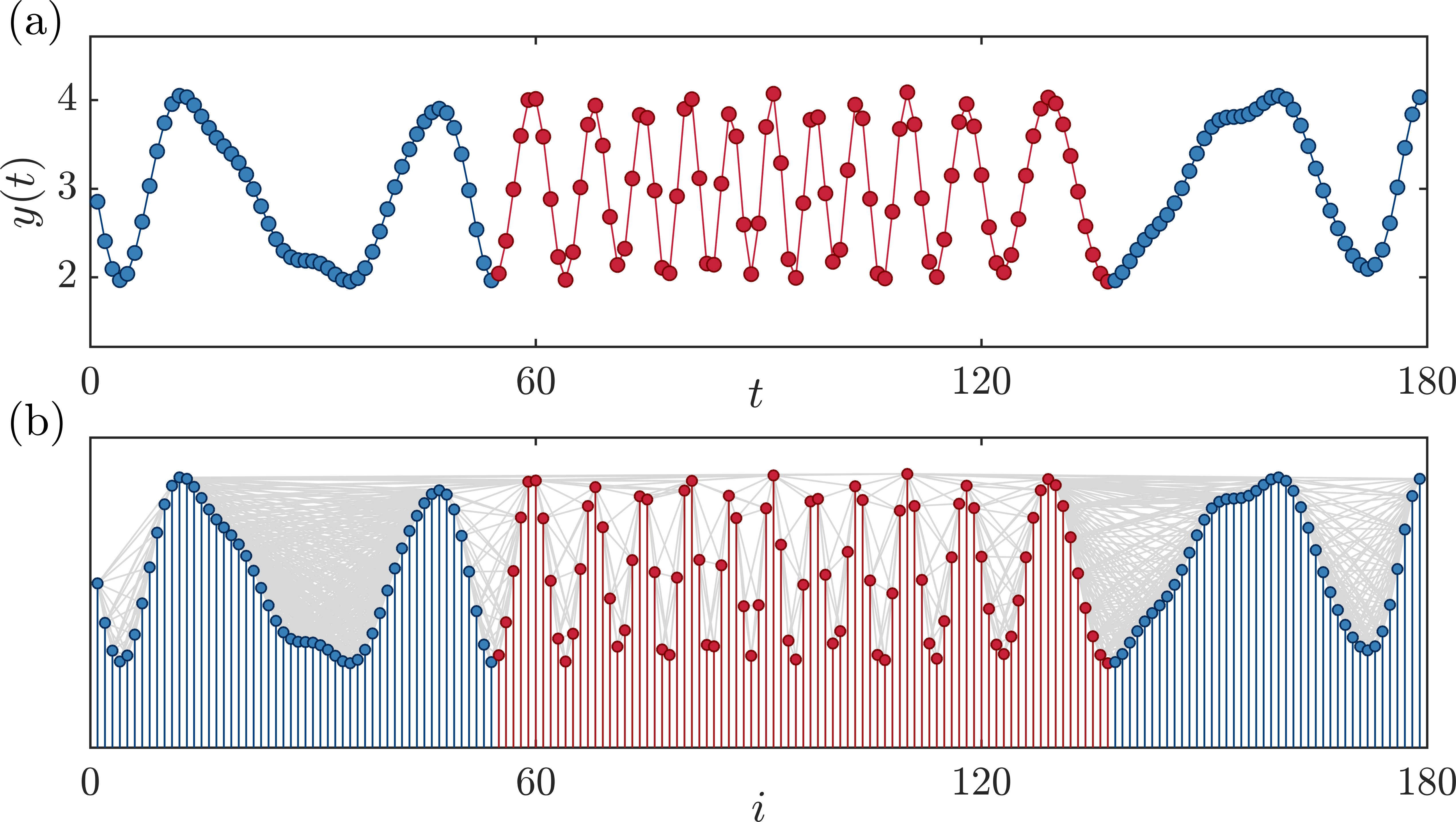}
    \caption{(a) A typical time series ($y(t)$) exhibiting frequency modulation and (b) its corresponding natural visibility graph (NVG) with time stamps as nodes. The blue and red stems at each time stamp indicate the amplitude of $y$ at that instant. Links between nodes are indicated by gray lines and determined by visibility algorithm. From the schematic it is evident that in the epochs of low-frequency dynamics (blue parts of the time series), the density of connections between nodes in the NVG is higher than that during epochs of high-frequency variations. Thus, the variations in the number of connections of a node in NVG captures the variation in frequency of a time series.}
    \label{fig_schematic}
\end{figure}

NVG captures the frequency modulation (FM) in an uniformly sampled time series \citep{iacobello2018visibility,iacobello2021large}. Consider a typical time series exhibiting FM and the corresponding NVG shown in figure \ref{fig_schematic}(a) and (b), respectively. Note that, the visibility of a node close to the trough is restricted by the neighboring peaks in the time series. For a time series with uniform sampling rate, one cycle of a high-frequency periodic oscillation has a smaller time period and contains lesser number of time stamps (nodes) compared to that of a low-frequency oscillation. For example, a node in a high-frequency oscillation cycle can connect to lesser number of nodes as compared to a node in low-frequency oscillation. Thus, the variation in the number of connections of a node quantifies the variation in the frequency of the time series. The number of connections of a node $i$ is called its degree, which is expressed as $k_i = \sum_{j = 1} ^{N} a_{ij}$.

Compare the number of connections between two consecutive peaks during low and high-frequency dynamics in figure \ref{fig_schematic}(b). Clearly, the average degree of a node in the high-frequency epoch is lesser than that of a node in the low-frequency epoch. On comparing the average degree of nodes during different epochs, one can infer the frequency modulation in the signal. \textcolor{black}{Recently, this feature of NVG was used to study frequency modulation in the stream-wise and span-wise components of velocity fluctuations in wall-bounded turbulence resulting from the interaction between small- and large-scale flow structures \citep{iacobello2021large}. Here, we apply this method \citep{iacobello2021large} to study interactions in the form of cross-variable amplitude-frequency coupling arising between the acoustic and heat release rate signals. }

\subsection{Cross-variable conditionally averaged degree from natural visibility graphs}\label{sec_NVGconditional}

From the spectrogram in figure \ref{fig_spectrogram}(b), we notice that the amplitude of the dominant frequency of the acoustic pressure signal is significant during the epochs of high-amplitude periodic oscillations as compared to that of low-amplitude fluctuations in $p'$. We identify epochs of high- and low-amplitude $p'$ dynamics using short-time Fourier transform performed on non-overlapping windows of the time series of $p'$. The size of a window $w$ corresponds to $L=5$ cycles of the dominant acoustic mode. We find the amplitude of the frequency of the dominant mode ($A_{DM}^w$) from the short-time Fourier transform for each window $w$. Then we set a threshold on these amplitudes of dominant frequency across all windows as $A_\tau= 15\%$ of $\text{max} (A_{DM}^w)$. Then, any window $w$ for which $A_{DM}^w>A_\tau$ is identified as a window of high-amplitude $p'$ dynamics, else as an epoch of low-amplitude $p'$ dynamics. In Appendix \ref{app_FFT} we show that the spatial pattern of cross-variable amplitude-frequency coupling obtained using this approach remains qualitatively similar for small variations in $L$ and $A_\tau$.  

In order to identify the spatial pockets of frequency modulation in heat release rate dynamics, we construct NVG of $\dot{q}'(x,y)$ fluctuations at each location in the flow field. Then, we find the conditionally averaged degree of nodes included in all those windows which are identified as epochs of high-amplitude (or low-amplitude) $p'$ dynamics using the criterion discussed above, and denote it as $K_h$ (or $K_l$) as given by equation \ref{eqn_KlKh}. Thus, 
\begin{equation}\label{eqn_KlKh}
\begin{split}
K_{h}(x,y) &= \frac{1}{N_h}\sum_{j=1}^{N} (k_j(x,y)|j\in w: A_{DM}^w>A_\tau)\\
K_{l}(x,y) &= \frac{1}{N_l}\sum_{j=1}^{N} (k_j(x,y)|j\in w: A_{DM}^w<A_\tau)
\end{split} 
\end{equation}
where $(x,y)$ denotes a spatial location, $N_h$ and $N_l$ are the total number of nodes (time-stamps) during the high- and low-amplitude epochs in the time series of $p'$, respectively. Also, $k_j$ is the degree of the $j^{th}$ node, $j\in w$ denotes that $w$ is the short-time window containing node $j$, and $N$ is the total number of nodes. Note that, $K_h$ and $K_l$ are conditional averages of degree of nodes in NVG of $\dot{q}'$ conditioned over the amplitude of acoustic pressure oscillations, and are thus called conditional average degree. 

Next, we compute the ratio $K_{lh}(x,y) = {K_l(x,y)}/{K_h(x,y)}$ as proposed by \citet{iacobello2021large}, to quantify changes in the frequency of $\dot{q}'$ at a location $(x,y)$ with variation in the amplitude of $p'$. Note that, \citet{iacobello2021large} have shown that the ratio of conditionally averaged degrees obtained from NVG reflects frequency modulation alone, even if both \textcolor{black}{amplitude and frequency modulation} occur in a signal.

From figure~\ref{fig_schematic}, we understand that the greater the average degree in a time-window, the lower the frequency of oscillations in that window. Then, if $K_{lh}(x,y)>1$, i.e., $K_l(x,y)>K_h(x,y)$, the frequency of $\dot{q}'$ fluctuations at location $(x,y)$ is lower during the epochs of low-amplitude $p'$ dynamics as compared to that for epochs of high-amplitude $p'$ dynamics. Thus, $K_{lh}(x,y)>1$ signifies that on an average the frequency of $\dot{q}'$ fluctuations at location $(x,y)$ increases when the amplitude of $p'$ oscillations increases, and we refer to this condition as positive frequency modulation (FM). In contrast, if $K_{lh}(x,y)<1$, we infer that on an average the frequency of $\dot{q}'$ fluctuations at location $(x,y)$ decreases when the amplitude of $p'$ oscillations increases, and we refer to this condition as negative FM.

The method using cross-variable conditionally averaged degree of NVG introduced here helps capture simultaneous co-evolution of and coupling between the frequency of one variable and amplitude of another. This approach is advantageous compared to conventional spectrogram analysis where frequency modulation in one variable is detected independent of other variables in the system.


\section{Results}\label{sec_result_all}

\subsection{Occurrence of intermittency during quasi-static experiments}\label{sec_result_quasi}

\begin{figure}
    \centering
    \includegraphics[width=1\linewidth]{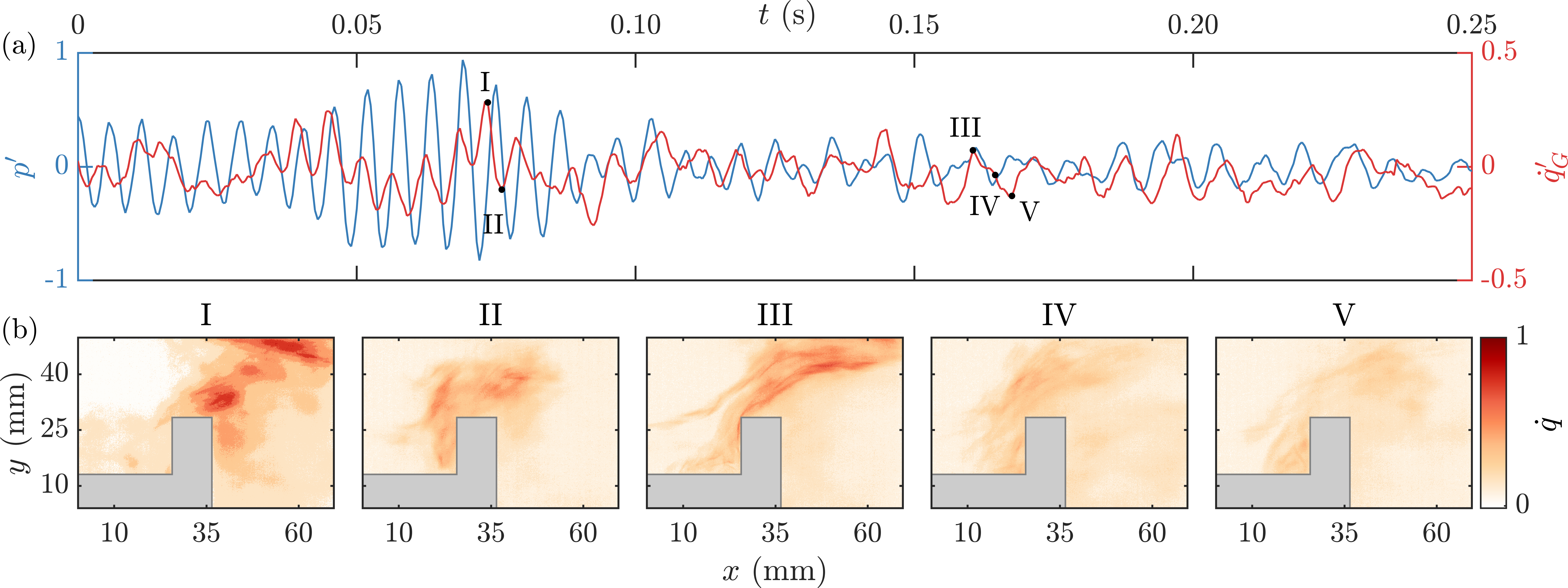}
    \caption{(a) Normalised time series of acoustic pressure ($p'$) and global heat release rate ($\dot{q}_G'$) fluctuations, during the state of intermittency observed in quasi-static experiments in a turbulent bluff-body stabilized combustor. (b) The spatial pattern of heat release rate ($\dot{q}(x,y)$) during high-amplitude (I-II) and low-amplitude (III-V) acoustic pressure dynamics corresponding to the points marked in the time series plot in (a). The strength and location of significant heat release are different during epochs of high- and low-amplitude amplitude $p'$ oscillations.}
    \label{fig_rawdata}
\end{figure}

In quasi-static experiments, we vary the flow control parameter such that fully developed dynamical states are achieved at each control parameter. As described in Section \ref{sec_expt}, we observe a dynamical transition in the acoustic pressure fluctuations from chaotic to periodic dynamics through a state of intermittency. Here, we analyse the cross-variable coupling between the acoustic pressure and the heat release rate fluctuations during the state of intermittency using the method described in Section \ref{sec_dataAnalysis}. Figure \ref{fig_rawdata}(a) shows the normalised time series of the acoustic pressure ($p'$) and the global heat release rate ($\dot{q}_G'$) fluctuations obtained during this state. Each time series is normalised by its maximum value. Figure  \ref{fig_rawdata}(b) shows the spatial pattern of the instantaneous heat release rate ($\dot{q}(x,y)$) corresponding to the points (I-V) marked in the time series of $\dot{q}_G'$ in figure \ref{fig_rawdata}(a). Distinct spatial patterns of heat release rate occur during high- and low-amplitude $p'$ dynamics. 

When $\dot{q}_G'$ is at a local maxima during the high-amplitude $p'$ dynamics (see figure \ref{fig_rawdata}(b-I)), the heat release rate is very high specifically in the inner shear layer and regions downstream of the bluff-body (ISL and W regions in figure \ref{fig_exptschematic}). When $\dot{q}_G'$ is at a local minima during the high-amplitude $p'$ dynamics (see figure \ref{fig_rawdata}(b-II)), the heat release rate is significant around the bluff-body, particularly in the inner shear layer and in the region of secondary flow recirculation (i.e., regions ISL and SFR as shown in figure \ref{fig_exptschematic}(b), respectively). On the other hand, when $\dot{q}_G'$ is maximum during the low-amplitude $p'$ dynamics (see figure \ref{fig_rawdata}(b-III,IV)), the heat release rate is significant in the ISL, OSL and SFR regions. Finally, when $\dot{q}_G'$ is minimum during the low-amplitude $p'$ dynamics, the strength of heat release rate reduces and is significant predominantly in SFR and ISL region (see figure \ref{fig_rawdata}(b-V)). 

\begin{figure}
    \centering
    \includegraphics[width=1\linewidth]{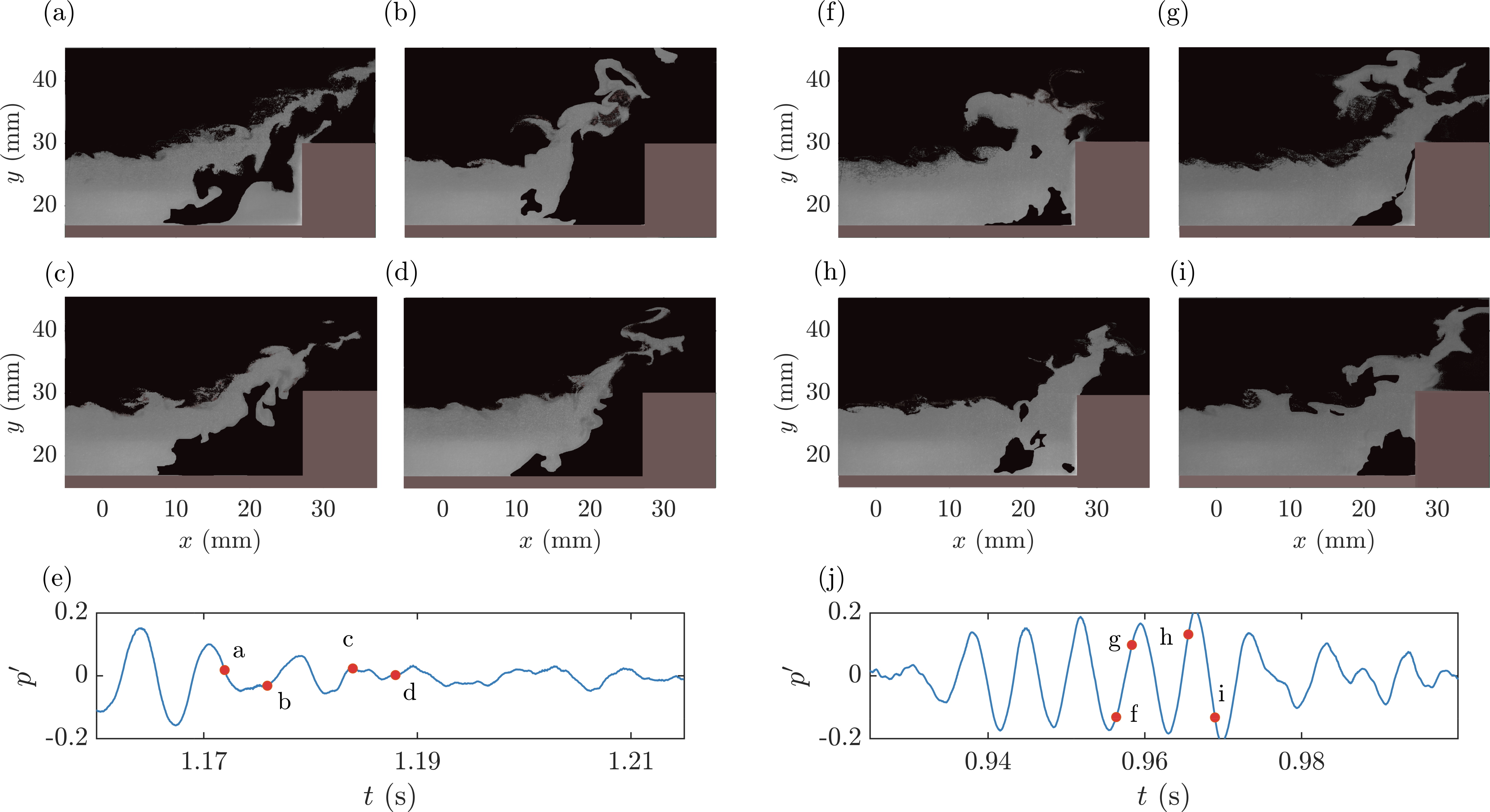}
    \caption{Oil Mie scattering images (a-d) and (f-i) delineating the flame anchoring locations and flame front during epochs of (e) aperiodic and (f) periodic fluctuations in $p'$, respectively, during the state of intermittency obtained in quasi-static experiments in a turbulent bluff-body stabilized combustor. The flame anchors predominantly in the dump plane at the edge of the outer shear layer. Along the shaft holding the bluff-body, the flame anchoring extends into the secondary flow recirculation region during periodic epochs but remains restricted upstream during aperiodic epochs of intermittency.}
    \label{fig_quasimie}
\end{figure}

Figure \ref{fig_quasimie} shows the oil Mie scattering images delineating the flame front and the anchoring position during the state of intermittency obtained during quasi-static experiments in the combustor. During epochs of aperiodic fluctuations in $p'$ (such as in figure \ref{fig_quasimie}(e)), the flame anchors predominantly at two locations, one in the OSL close to the backward-facing step and another along the shaft in the ISL region upstream of the bluff-body as evident in figure \ref{fig_quasimie}(a-d). The flame can also anchor in the SFR region, although rarely, during such aperiodic epochs (as in figure \ref{fig_quasimie}(a)). Further, small-scale distortions are evident along the flame front in the OSL, ISL and SFR regions (see figure \ref{fig_quasimie}(a-d)). 

During epochs of high-amplitude periodic oscillations in $p'$, the flame anchors in an extended region along the shaft. The flame
anchoring location in the OSL region close to the backward-facing step is similar during both periodic and aperiodic epochs (see figure \ref{fig_quasimie}(f-i)). However, the flame also anchors often in the SFR region during the epochs of periodic fluctuations in $p'$ (see figure \ref{fig_quasimie}(g,h)). The flame front in the OSL region exhibits large scale distortions induced by large scale coherent structures shed during this epoch (see figure \ref{fig_quasimie}(f,g,i)). Also, the flame anchors in the ISL region close to the shaft and the flame front in ISL region exhibits small scale distortions that merges with the flame anchored in the SFR region (see figure \ref{fig_quasimie}(f-h)). The turbulent flame front extends and oscillates between the OSL and recirculation zone in the dump plane (DR region) and the part of inner shear layer above and downstream of the bluff-body.

\begin{figure}
    \centering
    \includegraphics[width=\linewidth]{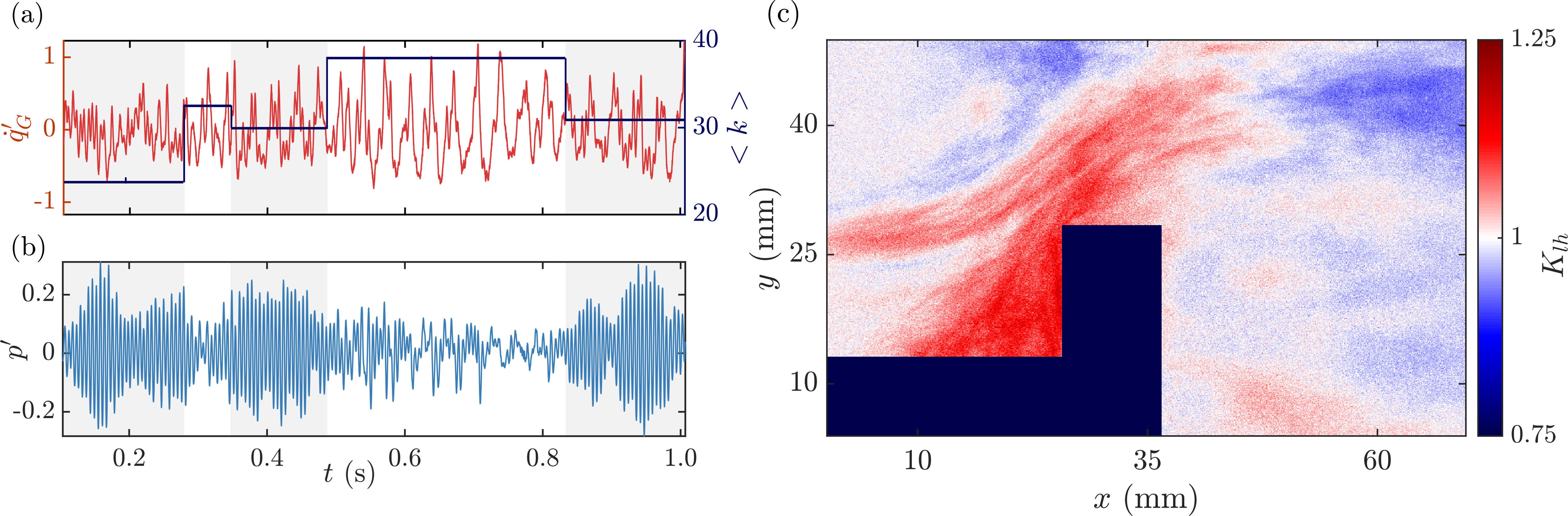}
    \caption{Time series of (a) global heat release rate fluctuations (${\dot{q}'_G}$, red curve), the variation of short-time averaged degree ($\langle k \rangle$, blue curve) obtained from the NVG of ${\dot{q}'_G}$, and (b) the  time series of acoustic pressure ($p'$). The shaded regions represent epochs of high-amplitude $p'$ dynamics. (c) Spatial distribution of the ratio of conditionally averaged degree during low- and high-amplitude $p'$ dynamics, $K_{lh}$, during the state of intermittency obtained during quasi-static experiments in a turbulent bluff-body stabilized combustor. The average degree $\langle k \rangle$ (in subplot (a)) is lower and hence the frequency of $\dot{q}_G'$ is higher during epochs of high-amplitude $p'$ oscillations as compared to that in low-amplitude $p'$ fluctuations. Spatially, $\dot{q}'$ exhibits positive frequency modulation specifically in the outer shear layer, parts of the inner shear layer and secondary flow recirculation zone, which are also the regions of flame anchoring, and negative frequency modulation in spatial pockets in the dump plane which are also the regions of significant flame front distortion.}
    \label{fig_quasiresult_KLH}
\end{figure}

Figure \ref{fig_quasiresult_KLH}(a) and (b) show the time series of global heat release rate ($\dot{q}_G'$) and acoustic pressure ($p'$) oscillations, respectively. The time series of $p'$ is partitioned into epochs of low- and high-amplitude oscillations using short-window fast Fourier transform as described in Section \ref{sec_NVGconditional}. The epochs of high-amplitude $p'$ dynamics are shaded gray in figure \ref{fig_quasiresult_KLH}(a,b). We construct natural visibility graphs (NVG) from the time series of global heat release rate fluctuations ($\dot{q}_G'$) as described in Section \ref{sec_NVG_construct}. Then, we compute the average degree ($\langle k \rangle$) over short time windows corresponding to five acoustic cycles as shown by the blue curve in figure \ref{fig_quasiresult_KLH}(a). Clearly, the value of $\langle k \rangle$ is relatively lower during epochs of high-amplitude $p'$ dynamics, as compared to the epochs of low-amplitude $p'$ dynamics. Remember, higher the average degree of NVG in an epoch, lower the frequency of oscillations in that epoch of the corresponding time series (Section \ref{sec_NVGconditional}). Thus, we infer that the frequency of $\dot{q}_G'$ is higher during epochs of high-amplitude $p'$ dynamics as compared to low-amplitude $p'$ dynamics.


Next, we construct NVG of local heat release rate fluctuations ($\dot{q}'(x,y,t)$) at each location $(x,y)$ in the combustor and compute $K_{lh}$, that is the ratio of short-time average conditional degree during low- and high-amplitude $p'$ dynamics (as discussed in Section  \ref{sec_NVGconditional}). Note that, $K_{lh}>1$ at a location implies positive frequency modulation (FM), that is, frequency of $\dot{q}'$ increases with the amplitude of $p'$. 
We find that positive FM occurs in the regions associated with flame anchoring during epochs of high-amplitude $p'$ dynamics, namely the OSL and SFR regions and part of the ISL region close to the shaft (compare red regions in figure \ref{fig_quasiresult_KLH}(c) with the flame in figure \ref{fig_quasimie}(f-i)). 

Also, negative FM ($K_{lh}<1$) occurs in a spatial pocket in the dump plane close to the OSL which is associated with large-scale flame front distortions during the epochs of high-amplitude $p'$ dynamics (compare blue pocket in the DR region in figure \ref{fig_quasiresult_KLH}(c) with figure \ref{fig_quasimie}(f,g)). Negative FM also occurs in pockets of the ISL and wake (W region) downstream of the bluff-body and close to the top wall of the combustor. There is no significant FM in $\dot{q}'$ and $K_{lh}\approx 1$ in the core of the turbulent flame in the ISL region very close to the shaft of the bluff-body.


In summary, we find that the $\dot{q}'$ fluctuations exhibit lower frequency in the regions associated with flame flapping and higher-frequency in the regions of flame anchoring during high-amplitude $p'$ dynamics as compared to low-amplitude $p'$ dynamics. Our finding implies that a unique spatial pattern of cross-variable amplitude-frequency coupling emerges during the state of intermittency. This pattern of interaction can be understood in terms of the physical processes, as following.

During the state of intermittency, high-amplitude periodic $p'$ oscillations are sustained for short epochs owing to flame-acoustic interactions. 
High-frequency fluctuations in $\dot{q}_G'$ during epochs of high-amplitude $p'$ dynamics (as detected in figure \ref{fig_quasiresult_KLH}(a)) imply that the aggregate heat release rate fluctuations are possibly more often in phase with the acoustic pressure fluctuations thus sustaining the periodic $p'$ dynamics. However, due to acoustic damping and turbulence in the flow and flame dynamics, high-amplitude $p'$ fluctuations are not sustained for very long. 

We find that, during epochs of high-amplitude $p'$ oscillations, large coherent structures form and shed in the dump plane and convect downstream from the top of the bluff-body into the wake \citep{george2018pattern}. The size and shedding frequency of these vortices depends on fluctuations in the acoustic field \citep{chu1958non, matveev2003model}. Large vortices delay mixing as they entrain and trap the unburnt reactants. Combustion occurs after the entrained reactants mix completely with hot products contained in the recirculation zones, and leads to sudden heat release. Sometimes these vortices can hit the bluff-body or the walls of the combustor and break into smaller vortices enhancing the mixing \citep{schadow1992combustion}. The large vortices are shed almost periodically and distort the flame as they convect downstream during epochs of high-amplitude $p'$ dynamics \citep{george2018pattern}. 

Larger the vortex, larger the scale of flame front distortion. Such flame front distortion causes flapping of the free end of the flame and induces low-frequency oscillations in the $\dot{q}'$ dynamics during these epochs. In contrast, during low-amplitude $p'$ fluctuations, the smaller vortices introduce small scale distortions in the flame inducing comparatively higher-frequency fluctuations in the free end of the flame. As a result, we find  negative FM in local heat release in regions associated with flame front distortion and flapping, that is, in the spatial pockets in the dump plane and in the wake region close to the top wall of the combustor (blue pockets in figure \ref{fig_quasiresult_KLH}(c)). 


On the other hand, the large coherent structures discussed above do not perturb the flow and flame in the ISL and SFR regions close to the bluff-body shaft and in the OSL region close to the backward-facing step. Hence, these are the regions of flame anchoring during high-amplitude periodic $p'$ oscillations during the state of intermittency. When the amplitude of $p'$ oscillations is high, the acoustic field perturbations induce stronger excitation to the flow and flame dynamics. As a result, the turbulent flow and flame dynamics are excited at higher frequencies during epochs of high-amplitude $p'$ compared to low-amplitude $p'$ dynamics in these regions. Hence, we find positive FM in the regions where the flame anchors during high-amplitude $p'$ dynamics, that is the OSL, ISL and SFR regions.

\textcolor{black}{The frequency of local heat release rate fluctuations varies in time differently at distinct locations. As a result, the phase between the acoustic pressure and local heat release rate fluctuations shifts dynamically and alters the spatial pattern of thermoacoustic power generation. In Appendix \ref{app_RI_vortexsizedist}, using the spatial variation of Rayleigh index (a measure of thermoacoustic power), we show that the locations of positive FM are associated with stronger sources or weaker sinks of thermoacoustic power during high-amplitude $p'$ dynamics.}

We note that, not only does the coupling between $p'$ and $\dot{q}'$ dynamics determine the system dynamics, but the state of the system also determines the strength and pattern of these interactions. During the state of thermoacoustic instability, the effect of system damping and turbulence is dominated by the nonlinear feedback between acoustic, combustion and fluid dynamics leading to a self-sustained high-amplitude acoustic pressure dynamics. Evidently, such high-amplitude $p'$ oscillations are associated with self-sustained high-frequency oscillations in the heat release rate dynamics \citep{george2018pattern}. Our analysis during the state of intermittency reveals a spatial pattern of positive and negative frequency modulations with the amplitude of $p'$ and thus highlights the interactions that are possibly responsible for frequency shift in $\dot{q}'$ as the system dynamics transitions from chaos to order as reported by \citet{pawar2017thermoacoustic}.

\subsection{Occurrence of intermittency during continuous variation of flow control parameter}\label{sec_result_rate}

We perform a set of experiments in a turbulent bluff-body stabilized combustor where the air flow rate is varied continuously at a finite rate as described in Section \ref{sec_expt}. As the inlet air flow rate is varied at a finite rate, the Reynolds number ($Re$) of the inlet flow increases linearly while the equivalence ratio ($\phi$) decreases linearly. The time series of acoustic pressure fluctuations ($p'$), the global heat release rate ($\dot{q}_G$) and the spatial pattern of instantaneous heat release rate ($\dot{q}(x,y)$) are shown in figure \ref{fig_ratedata_raw}. The time series of $p'$ and $\dot{q}_G$ exhibit a dynamical transition from the state of intermittency to self-sustained limit cycle oscillations as $Re$ increases continuously. Also, from figure \ref{fig_ratedata_raw}(a,b), we observe that the average amplitude of $p'$ and $\dot{q}_G$ oscillations increase continuously with monotonic increase in $Re$. 

We decipher the relation between the flame and acoustic field dynamics during the dynamical transition induced by a continuous variation in the control parameter.  To do so, we divide the entire data into short time-windows of $T_w=0.6$ s. Note that, the average increase in the amplitude of $p'$ within one time frame $T_w$ is negligible compared to the amplitude variation between periodic and aperiodic fluctuations in $p'$ during that window (see insets A-C in figure \ref{fig_ratedata_raw}). Hence, we analyse each of these time-windows similar to that done for quasi-static experiments in the previous subsection. For each time-frame ($T_w=0.6$ s is equivalent to $68$ cycles of the dominant acoustic mode), we compute the short-window fast Fourier transform over a smaller time period corresponding to five cycles of the dominant acoustic mode ($L=5$ as described in Section \ref{sec_NVGconditional}). We then classify the epochs of high- and low-amplitude $p'$ dynamics within each time frame using the method described in Section \ref{sec_NVGconditional}.

Figure \ref{fig_rate_results} shows the spatial distribution of $K_{lh}$ during consecutive time-windows (each of $0.6$ s). Here, we classify the dynamical transition observed in experiments into three regimes, namely, low, moderate and high amplitude intermittency dynamics (for example A, B, C-inset in figure \ref{fig_ratedata_raw}(a,b), respectively) that occur prior to the onset of high-amplitude self-sustained periodic oscillations. We observe that the acoustic pressure, global and local heat release rate dynamics are distinct during these regimes. To demarcate these dynamical regimes quantitatively, we use the variations in $p'_{rms}$ and $\dot{q}_{G_{rms}}$ calculated over each time-window of $T_w=0.6$ s as discussed in Appendix \ref{app_classifyregimes}. In Appendix \ref{app_Tw}, we show that the flame-acoustic interactions inferred during each dynamical regime are not sensitive to small changes in the value of $T_w$. 

\subsubsection*{(A) Low-amplitude intermittency regime:} 
\begin{figure}
    \centering
    \includegraphics[width=1\linewidth]{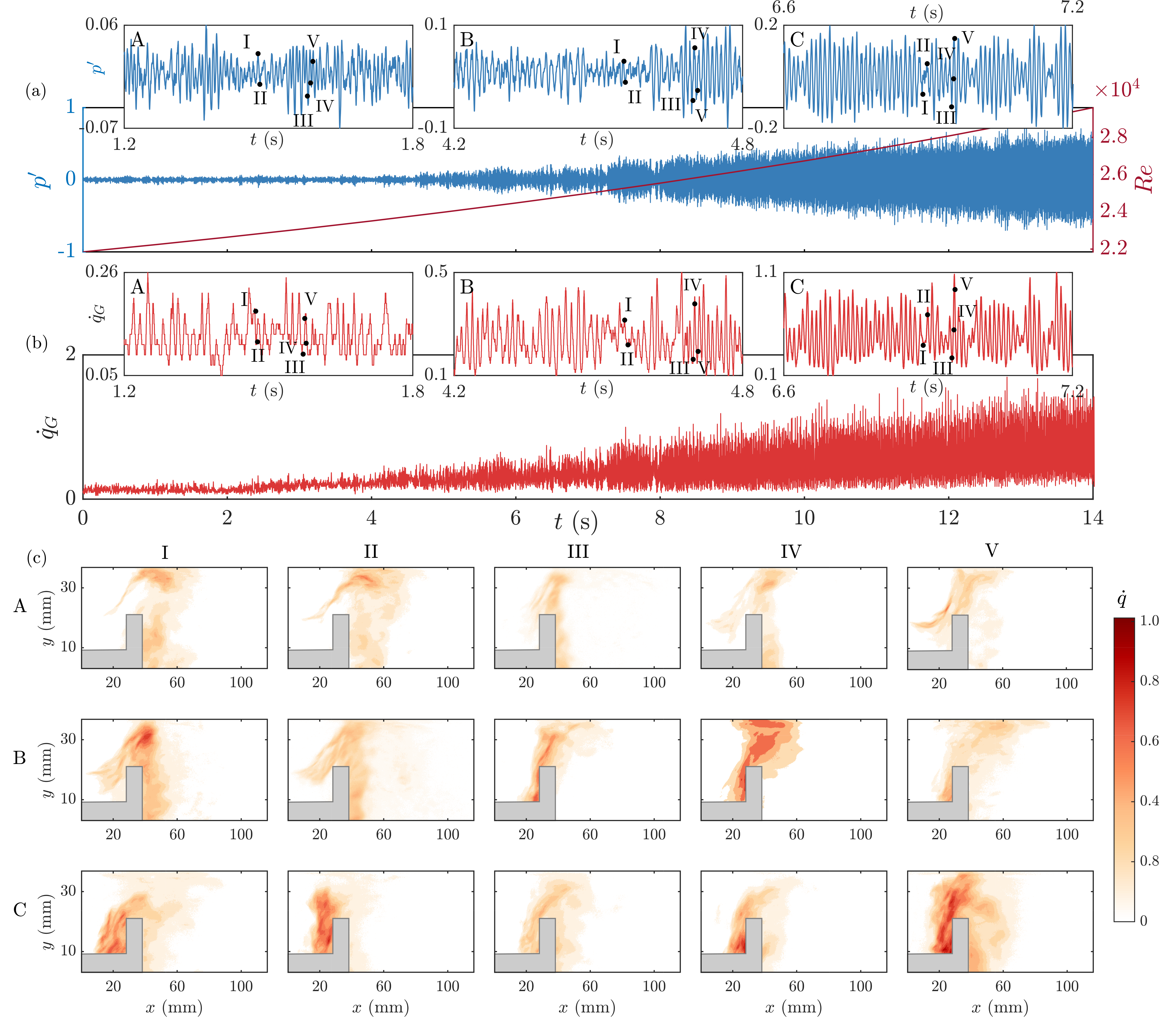}
    \caption{(a) Time series of  acoustic pressure fluctuations ($p'$, blue curve), Reynolds number of inlet fuel-air mixture ($Re$, red line) and (b) global heat release rate ($\dot{q}_G$) obtained from experiments in a turbulent bluff-body stabilized turbulent combustor when the inlet mass flow rate of air ($\dot{m}_a$) is varied at a finite rate while the mass flow rate of fuel ($\dot{m}_f$) is maintained at a constant value. (c) The spatial heat release rate ($\dot{q}(x,y)$) observed in the combustor shown in rows (A-C) corresponding to each of the insets (A-C) in the time series of $\dot{q}_G$ in (b). The labels (I-V) for each row correspond to points (I-V) labeled in the time series in each of the insets in (b). The rows A, B, C represent the spatial heat release pattern during the occurrence of low, moderate and high-amplitude intermittency, respectively. The average strength of heat release rate increases with $Re$ as evident in the time series of $\dot{q}_G$ as well as spatial distribution of $\dot{q}$. Further, the spatial pattern of significant heat release varies significantly with the average increase in $p'$ as well as with amplitude variations during intermittent oscillations in $p'$ during short time-windows.} 
    \label{fig_ratedata_raw}
\end{figure}

At low $Re$, the time series of $p'$ and $\dot{q}_G$ exhibit low-amplitude intermittency dynamics with very low-amplitude aperiodic fluctuations interspersed between short epochs of relatively high-amplitude periodic oscillations (see inset A in figure \ref{fig_ratedata_raw}(a,b)). 
During our experiments, we observe that the flame is anchored predominantly in the OSL region close to the backward-facing step. The free end of the turbulent flame exhibits significant oscillations between the OSL and ISL regions and the dump plane (DR region). Thus, we find that significant heat release occurs predominantly in the ISL region above and downstream of the bluff-body and in the OSL (see figure \ref{fig_ratedata_raw}(c)-A(I-V) and figure \ref{fig_app_meanHRRrate}(I-IV) in Appendix \ref{app_rate_meanHRR}).

The spatial distribution of $K_{lh}$ obtained during such low-amplitude intermittency dynamics is shown in figure \ref{fig_rate_results}(I-IV). We find that positive FM ($K_{lh}>1$) occurs predominantly in the OSL and also in a spatial pocket in the dump plane (see figure \ref{fig_rate_results}(II,IV)). Further, $K_{lh}>1$ in parts of the ISL above the bluff-body and also in the wake downstream of and very close to the bluff-body (see figure \ref{fig_rate_results}(I,III,IV)). In summary, positive FM occurs predominantly in the recirculation zones in the dump plane and wake of the bluff body (DR and W regions) and also in the location of flame anchoring, that is the OSL. 

On the other hand, $K_{lh}<1$ (signifying negative FM) in certain pockets in the dump plane aligned adjacent to the OSL (see figure \ref{fig_rate_results}(II,IV)) and in pockets in the wake close to the top wall of the combustor (see figure \ref{fig_rate_results}(I,III)). These regions are associated with the free oscillating end of the flame and negative FM is induced due to flame distortion by coherent structures. Note that, the coherent structures shed are relatively smaller in size and the epochs of periodic dynamics are relatively shorter during low-amplitude intermittency dynamics compared to other dynamical regimes. As a result, the flame distortion by vortices occurs in a small spatial extent and for shorter epochs. Hence, we find that negative FM occurs in small pockets associated with flame distortion that occurs typically in the dump plane and close to the top wall of the combustor downstream of the bluff-body. 

\subsubsection*{(B) Moderate-amplitude intermittency regime:}

Insets B in figure \ref{fig_ratedata_raw}(a, b) show the time series of  $p'$ and $\dot{q}_G$ during moderate-amplitude intermittency dynamics that occurs when $Re$ of the inlet flow increases continuously in experiments. During our experiments, we observe that the flame anchors in the dump plane and OSL close to the backward-facing step and also along the shaft of the bluff-body. Close to the shaft, the flame anchors primarily in the ISL region upstream of the SFR region and the flame front exists primarily in the OSL, ISL and SFR regions. However, sometimes the flame may also anchor in the SFR region and then the flame front oscillates in the ISL and W regions above and close to the bluff-body. 

We observe that when $p'$ exhibits relatively high-amplitude periodic dynamics, instantaneous heat release rate is most significant in the SFR region and ISL above the bluff-body (see figure \ref{fig_ratedata_raw}(c)-B(III-V)). On the other hand, when $p'$ exhibits relatively low-amplitude aperiodic fluctuations, significant heat release occurs in the SFR, OSL and ISL regions (see figure \ref{fig_ratedata_raw}(c)-B(I,II)). Detailed inferences about the flame dynamics are presented using the spatial pattern of mean heat release rate in Appendix \ref{app_rate_meanHRR}.

\begin{figure}
    \centering
    \includegraphics[width=1\linewidth]{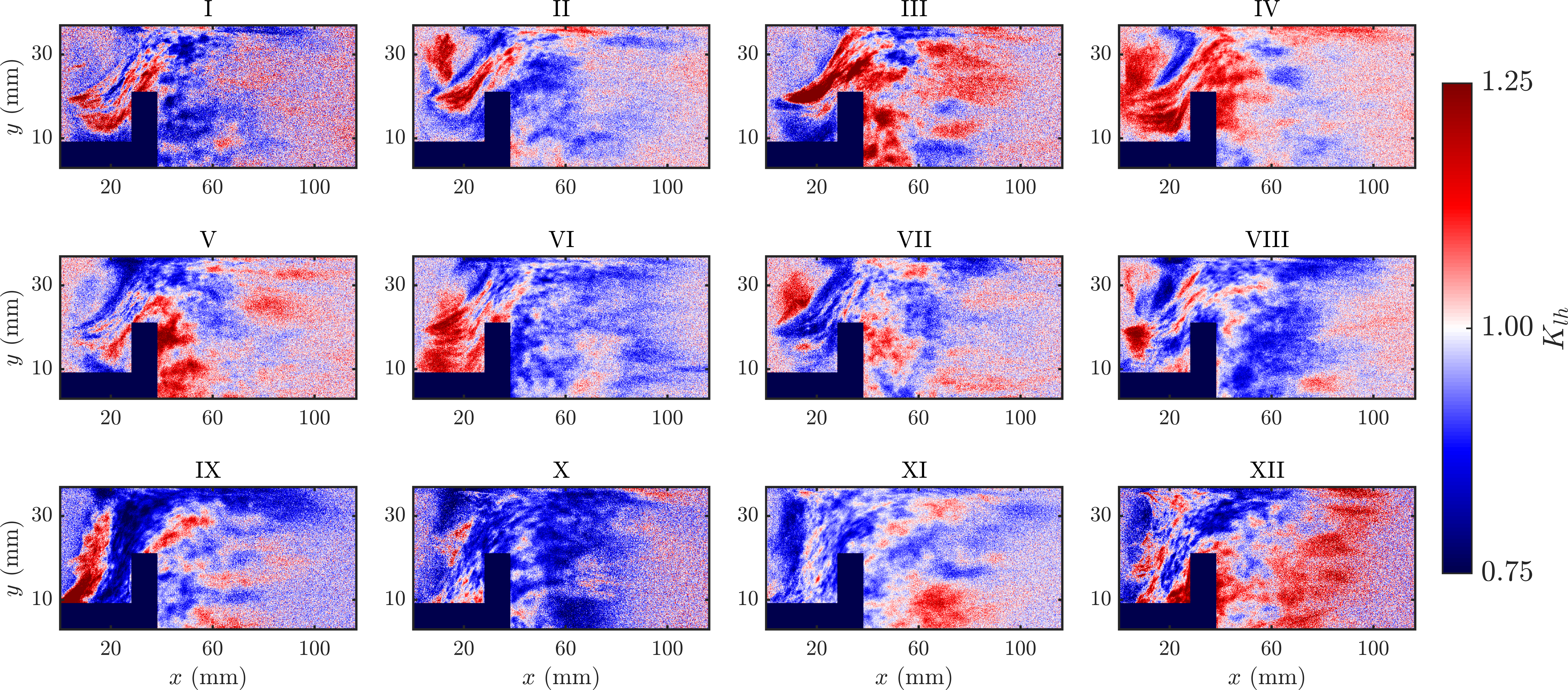}
    \caption{Spatial distribution of the ratio of conditional degree during low- and high-amplitude $p'$ dynamics, $K_{lh}$, obtained from the NVG of $\dot{q}'$ dynamics for consecutive time-windows (each of $T_w=0.6$ s) during dynamical transition in a turbulent bluff-body stabilized combustor when the air flow rate is varied at a finite rate. Here, (I-IV), (V-VIII) and (IX-XII) correspond to the epochs when the average amplitude of acoustic pressure oscillations is low, moderate and high during the occurrence of intermittency, respectively. The classification of the average amplitude into these three ranges is made qualitatively. Regions of strong positive and negative frequency modulation are evident during the occurrence of intermittency dynamics in experiments involving continuous variation of the airflow rate. A remarkable shift in such regions of positive and negative frequency modulation (FM) is observed with increase in the average amplitude of $p'$ during the state of intermittency. Variations in regions of positive and negative FM occur due to dynamically changing locations of flame anchoring and flame front distortions.}
    \label{fig_rate_results}
\end{figure}

Windows V-VIII in figure \ref{fig_rate_results} show the spatial distribution of $K_{lh}$ obtained during moderate-amplitude intermittency dynamics. Spatial pockets in the ISL close to the top wall of the combustor exhibit negative FM (figure \ref{fig_rate_results}(V-VIII)). Intriguingly, we find that negative FM occurs in the OSL in figure \ref{fig_rate_results}(V, VII) but positive FM occurs in spatial pockets in the OSL in figure \ref{fig_rate_results}(VI,VIII). In figure \ref{fig_rate_results}(VI), we find that $K_{lh}>1$ along the shaft, whereas in figure \ref{fig_rate_results}(V,VII) $K_{lh}\approx 1$ in the SFR region close to the shaft. Also, SFR region exhibits positive FM for some short epochs such as in figure \ref{fig_rate_results}(VIII). 

\textcolor{black}{We conjecture that, such alternating pattern of positive and negative FM occurs due to the appearance of large coherent structures and variations in flame anchoring position observed in experiments. Compared to the low-amplitude intermittency regime, epochs of periodicity are longer, the vortices shed in the dump plane are relatively larger, the mean flow velocity is greater and the typical local flow timescale is smaller during moderate-amplitude intermittency regime. Thus, the flame is able to anchor in the dump plane only for short epochs and tries to anchor in SFR region during other epochs.} Further, flame front distortion occurs predominantly in the OSL region or in some spatial pockets adjacent to OSL in the dump plane. The flame distortion extends well into the OSL region owing to the larger size vortices formed in the dump plane during this dynamical regime. 

As a result, we find negative FM in the OSL during some time-windows (figure \ref{fig_rate_results}(V,VII)). On the other hand, when the flame anchors close to the backward facing step, we obtain patches of positive FM in the OSL as shown in figure \ref{fig_rate_results}(V,VII). Further, the flame anchors not only in the OSL, but also in the SFR region as the flame is perturbed by relatively large vortices shed in the DR and OSL region. As a result positive FM is observed in the ISL and SFR regions close to the shaft such as in figure \ref{fig_rate_results}(VI,VIII). Furthermore, flame front oscillations are prominent in the ISL region above and downstream of the bluff-body. Hence, we find negative FM in these regions. The effect of large coherent structures on the flame anchoring and flame front distortion becomes more prominent during high-amplitude intermittency dynamics as discussed below.

\subsubsection*{(C) High-amplitude intermittency regime:}

With further continuous increase in $Re$, we observe high-amplitude intermittency dynamics characterized by long epochs of high-amplitude periodic oscillations interspersed with short epochs of low-amplitude aperiodic fluctuations in $p'$. We observe in our experiments that, the flame anchors primarily along the shaft of the bluff-body and not close to the backward-facing step anymore. Moreover the flame anchoring location keeps shifting dynamically along the shaft. Such shifts in the flame anchoring location along the shaft appear to occur more frequently during the high-amplitude (compared to moderate-amplitude) intermittency regime. The flame anchors either upstream of the SFR or in the SFR region close to the shaft. We obtain very high heat release rate in the OSL, ISL and SFR regions (refer figure \ref{fig_ratedata_raw}(c)-C(I-V)). Also, the turbulent flame extends downstream of the bluff-body and significant heat release occurs in the wake region (evident in figure \ref{fig_ratedata_raw}(C-V)). The flame dynamics during this dynamical regime is discussed further with respect to mean heat release rate during short time frames in Appendix \ref{app_rate_meanHRR}.

The spatial distribution of $K_{lh}$ obtained during the occurrence of high-amplitude intermittency is shown in figure \ref{fig_rate_results}(IX-XII). The pattern of frequency modulation is distinct in each time-frame. In figure \ref{fig_rate_results}(IX) we find  $K_{lh}>1$ in a spatial pocket in the ISL very close to the shaft of the bluff-body extending upwards into the OSL region. Moreover, we find  $K_{lh}<1$ in the SFR region and in the region above the bluff-body in figure \ref{fig_rate_results}(IX). Further, in figure \ref{fig_rate_results}(X,XI), we do not find any coherent spatial pockets exhibiting positive FM. Instead we find $K_{lh}<1$ in most regions in the flow field. Specifically, negative FM occurs in the SFR, OSL and ISL regions and in a large spatial pocket spanning across the recirculation zone in the dump plane (DR region). Finally, in figure \ref{fig_rate_results}(XII), we find negative FM occurs predominantly in the DR and OSL regions upstream and the ISL and W regions above and downstream of the bluff-body. We also find positive FM in the SFR region during the time-frame shown in figure \ref{fig_rate_results}(XII). Furthermore, incoherent patches of both positive and negative FM occur in very small pockets at scattered locations in the OSL and ISL region upstream of the bluff-body, evident in figure \ref{fig_rate_results}(X-XII).


Note that, the location of flame anchoring is stable if the mean flame front emerges from and remains attached to this location and if the flow dynamics entrains unburnt fuel alongside hot products across the flame front. However, as vortices form and convect downstream, the mixing of reactants and hot products is delayed and the primary location of recirculation changes dynamically with the convection of each large coherent structure. 
\textcolor{black}{As the mean flow velocity increases owing to continuous variations in the inlet airflow rate, much larger vortices are shed almost periodically in the dump plane and the frequency of local velocity fluctuations increases (that is typical flow timescale decreases). Moreover, the flame-acoustic interactions become stronger as the amplitude of acoustic pressure oscillations increase. Very high-amplitude $p'$ oscillations also imply strong acoustic velocity fluctuations which perturb the flame along with the turbulent flow dynamics. 
The flame anchors in the SFR region during short epochs due to the presence of very large vortex that forms in the dump plane and the OSL region. High-frequency fluctuations in the local velocity perturb the flame anchoring in SFR region. Thus, when the large vortex convects downstream or dissipates and until the next one appears, the flame tries to anchor upstream of the SFR region along the shaft. Hence, the flame anchoring location appears to oscillate to and fro along the shaft during this dynamical regime.} The flame distortion occurs in a much larger spatial extent during such epochs of high-amplitude $p'$ dynamics. 

Hence, large spatial pockets in the OSL and DR regions exhibit negative FM (such as in figure \ref{fig_rate_results}(X-XII)) due to low-frequency oscillations induced in $\dot{q}'$ by the large-scale flame front distortions. Further, we observe that negative FM occurs in regions close to the top wall of the combustor and downstream of the bluff-body (see figure \ref{fig_rate_results}(IX-XII)). As the large vortex convects downstream and hits the top wall of the combustor it dissociates into relatively smaller vortices which further distort the flame in the wake of the bluff-body. Such flame distortions induce negative FM in these downstream locations. Since the flame anchors along the shaft, one might expect positive FM in most regions along the shaft. However, the location of flame anchoring keeps oscillating to and fro along the shaft. Evidently, our method identifies the prominent location of flame anchoring during certain time-windows such as figure \ref{fig_rate_results}(IX,XII) since positive FM is associated with flame anchoring locations. 

When the flame anchors for comparatively longer epochs in the SFR region during a time-frame (of $0.6$ s), we obtain positive FM in a small pocket in the SFR region (see figure \ref{fig_rate_results}(XII)). On the other hand, when the flame anchors mostly upstream of the SFR region in a certain time-frame, we observe positive FM as seen in figure \ref{fig_rate_results}(IX). However, due to continual shifting of the flame anchoring location along the shaft during certain epochs, a low-frequency component is introduced in the local heat release rate oscillations. Also, due to shifts in the flame anchoring location, the SFR region experiences flame front oscillations for certain epochs which induce low-frequency fluctuations in heat release rate during high-amplitude $p'$ dynamics. Hence, we obtain negative FM in the SFR and ISL regions close to the shaft (see figure \ref{fig_rate_results}(IX-XI)). 



Finally we note that, a very large flame front distortion can also comprise small flame front distortions within it owing to the interaction of the flame front with coherent structures of multiple length scales in the underlying turbulent flow and with the acoustic field fluctuations. A detailed quantification of such small scale structures in the flame front within large scale flame distortion was done by \citet{raghunathan2020multifractal} using multifractal analysis. They also showed quantitatively that the effect on dynamics of the small scale structures is more dominant than that of the large scale coherent structures. We conjecture that incoherent patches of positive and negative FM appearing in the ISL and OSL regions are due to the multi-scale flame front distortion inducing spatially incoherent frequency modulations in the local heat release rate dynamics.



\begin{table}
  \begin{center}
\def~{\hphantom{0}}
\begin{tabular}{>{\centering\arraybackslash}m{2.3cm}>{\centering\arraybackslash}m{4.8cm}>{\centering\arraybackslash}m{5cm}}

\textbf{Dynamical regime}& \textbf{Result of cross-variable conditional NVG analysis} & \textbf{Physical reasoning}                                                                  
\\\hline
\textbf{Low-amplitude intermittency regime} & Positive FM in OSL, pockets in DR and W regions; negative FM in pockets in DR close to OSL and regions close to top wall of combustor  & Small vortices cause flame distortion in small spatial pockets in DR; flame anchoring is stable in OSL         
\\ \hline
\textbf{Moderate-amplitude intermittency regime} & ISL, SFR regions close to the shaft and OSL exhibit alternating pattern of positive and negative FM & Flame anchors primarily in OSL and along the shaft, anchoring location shifts to SFR sometimes; flame front distortion by larger vortices extends into OSL 
\\ \hline
\textbf{High-amplitude intermittency regime}& Positive FM either upstream of or in SFR in some epochs; negative FM in most of the flow field, also sometimes in location of flame anchoring along the shaft & Very large coherent structures cause very large flame front distortion in DR and OSL and continual oscillations in flame anchoring location along the shaft 
\\ \hline
\end{tabular}
\caption{Summary of results for different dynamical regimes obtained during the occurrence of intermittency prior to the onset of thermoacoustic instability when the air flow rate is varied continuously at a finite rate in experiments in a turbulent thermoacoustic system.}
\label{summarytable}
\end{center}
\end{table}


\section{Conclusion}

Dynamical transition from chaos to self-sustained order via the route of intermittency in turbulent thermoacoustic systems is well-known. In this work, we describe the rich spatial pattern of a unique cross-variable amplitude-frequency coupling between the co-evolving variables acoustic pressure ($p'$) and local heat release rate dynamics ($\dot{q}'$) in the system, during the state of intermittency obtained when the mass flow rate of the inlet air is varied either (a) quasi-statically and (b) continuously at a finite rate in a turbulent bluff-body stabilized combustor. Towards this purpose, we \textcolor{black}{apply} the method proposed by \citet{iacobello2021large} using natural visibility algorithm (NVG) to detect frequency modulation (FM). We \textcolor{black}{use} this method to study cross-variable amplitude-frequency coupling between the acoustic and heat release rate dynamics in our system. We construct visibility graph from the time series of $\dot{q}'$ at each location and compare the connectivity of nodes during epochs of low and high amplitude $p'$ dynamics. The epochs of low and high-amplitude $p'$ oscillations are distinguished using a threshold on the amplitude of the dominant frequency peak detected through short-window fast Fourier transform. \textcolor{black}{We note that this formalism helps us identify if amplitude modulations in $p'$ are associated with frequency modulations in $\dot{q}'$. Our analysis reveals the spatial patterns of simultaneous modulations in the topological properties (amplitude or frequency) \citep{estrada2023complex} of the system variables ($p'$ and $\dot{q}'$), and helps us infer the relation between physical processes in the system.}

We discover that the frequency of local heat release rate fluctuations increases in regions of flame anchoring and decreases in regions associated with flame front distortions, during bursts of high-amplitude $p'$ dynamics. The spatial pattern of such cross-variable amplitude-frequency coupling varies across different dynamical regimes of intermittency obtained when the inlet airflow rate is varied continuously (summarized in table \ref{summarytable}). As the mean flow velocity increases due to the continuous variation in the inlet airflow rate, the local flow timescale decreases, the mean acoustic pressure amplitude increases, epochs of periodicity in $p'$ increase and the mean size of coherent structures shed in the dump plane increase. As a result, the pattern of flow-flame-acoustic interactions vary and alter the flame anchoring location and characteristics of flame front distortion during low, moderate and high-amplitude intermittency dynamics. The understanding developed from such analysis can be useful in design and control of flame-acoustic interactions to increase operational margins of turbulent combustors. Further, the nature of flame-acoustic coupling can be investigated using this approach in other types of combustors with different flame holding mechanisms, for premixed or partially premixed flames, and during different routes of transition to thermoacoustic instability such as a continuous or abrupt transition. An interesting pursuit for future works is to study the variations in the spatial pattern of FM during experiments involving continuous variation of the flow control parameter when the control parameter is varied at faster or slower rates. 


\backsection[Acknowledgements:]{We acknowledge Dr. Vishnu R. Unni for fruitful discussions. This work was funded by the Office of Naval Research Global (ONRG) (grant number N629092212011). T. S. and R. R. acknowledge the support from Prime Minister Research Fellowship, Government of India.}

\backsection[Author contributions:]{S.T.: Writing (lead), analysis (equal), interpretation (lead), editing (equal), conceptualization (equal); A.B.: Writing (equal), analysis (lead), interpretation (equal), editing (equal), conceptualization (equal); R.R.: Experiments (lead), editing (equal); M.R.: Experiments (equal), editing (equal); G.C.: analysis (equal), interpretation (equal), editing (equal), conceptualization (equal); R.I.S.: Editing (lead), analysis (equal), conceptualization (equal), supervision (lead), funding (lead). The authors have no conflict of interest.}
\backsection[Declaration of Interests:]{The authors report no conflict of interest.}
\backsection[Data availability:]{The data that support the findings of this study are available from the corresponding author upon reasonable request.}


\appendix

\section{Effect of parameters used for short-window FFT to identify epochs of high- and low-amplitude dynamics in $p'$}\label{app_FFT}

In this work, we investigate the frequency modulation (FM) in local heat release rate ($\dot{q}'$) fluctuations associated with variations in the amplitude of acoustic pressure fluctuations ($p'$). Thus, we need to distinguish the epochs of high and low amplitude dynamics in $p'$. Towards this purpose we use short-window fast Fourier transform computed over a window size $w$ corresponding to $L$ cycles of the dominant acoustic mode. Also, we place a threshold $A_\tau$ on the amplitude of the dominant frequency obtained across all windows as described in section \ref{sec_NVGconditional}. In the analysis presented in the main text, we use $L=5$ and $A_\tau=15\%$.

\begin{figure}
    \centering
    \includegraphics[width=1\linewidth]{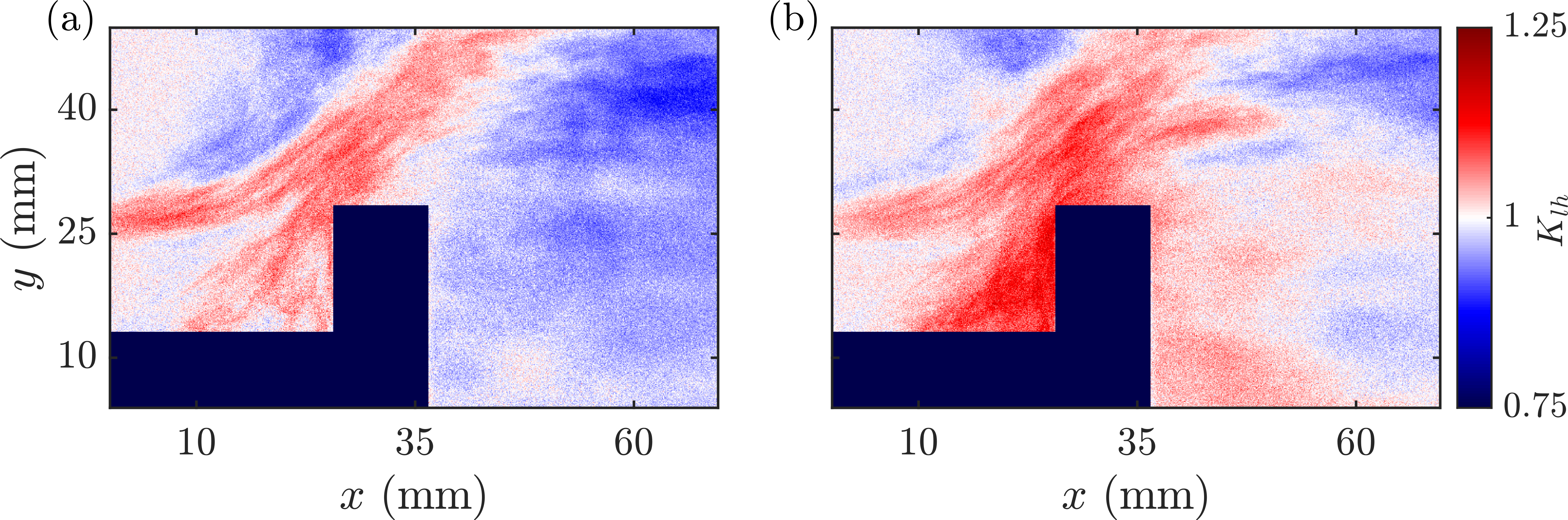}
    \caption{The spatial pattern of $K_{lh}$ obtained during the state of intermittency for quasi-static experiments when the short-window fast Fourier transform is computed for a window size $w$ corresponding to $L=5$ cycles of the dominant acoustic mode and the threshold on the amplitude of the dominant frequency is (a) $A_\tau=10 \%$ and (b) $A_\tau=20 \%$. The spatial pattern of FM in $\dot{q}'$ remains qualitatively the same for different $A_\tau$.}
    \label{fig_app_atau}
\end{figure}

Here, we examine the effect of changes in the parameters $L$ and $A_\tau$ used in the short-window FFT to identify epochs of high-amplitude $p'$ dynamics on the pattern of cross-variable amplitude-frequency coupling. From figure \ref{fig_app_atau}, we observe that on changing $A_\tau$, the threshold on the amplitude of dominant frequency obtained during short window FFT of $p'$, the spatial pattern of FM remains same qualitatively. However, the value and range of $K_{lh}$ changes slightly at each location.

\begin{figure}
    \centering
    \includegraphics[width=1\linewidth]{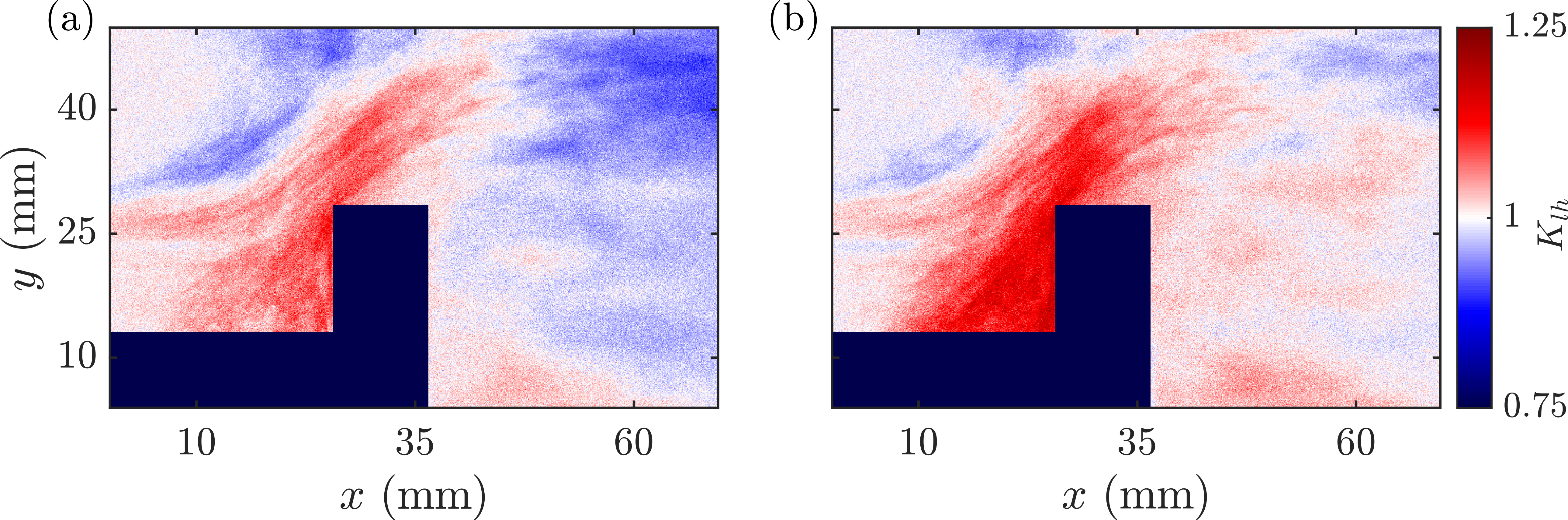}
    \caption{The spatial pattern of $K_{lh}$ obtained during the state of intermittency for quasi-static experiments when the short-window fast Fourier transform is computed for a window size $w$ corresponding to (a) $L=4$ and (b) $L=7$ cycles of the dominant acoustic mode. In each case, the threshold on the amplitude of the dominant frequency is $A_\tau=15\%$. The spatial pattern of FM in $\dot{q}'$ remains qualitatively the same for different sizes of the window used for short-window FFT.}
    \label{fig_app_L}
\end{figure}

Further, if we change the window size for which short window FFT of $p'$ is computed, we find again that the spatial pattern of FM is similar qualitatively (see figure \ref{fig_app_L}). Positive FM appears in regions of flame anchoring and negative FM occurs in regions of flame front oscillations. Negative FM (blue regions) in the dump plane is more pronounced in figure \ref{fig_app_L}(a) for shorter window sizes. Since the flame anchoring conditions do not vary significantly for long epochs, we see that positive FM in the secondary flow regions is more pronounced for larger window sizes as evident in figure \ref{fig_app_L}(b).

In summary, the results obtained for FM in local heat release rate fluctuations during high-amplitude $p'$ dynamics are not very sensitive to changes in the parameters of the short-window FFT used to identify epochs of high and low amplitude $p'$ fluctuations. 

\section{Spatial variation of Rayleigh index during high and low amplitude $p'$ dynamics}\label{app_RI_vortexsizedist}

\begin{figure}
    \centering
    \includegraphics[width=1\linewidth]{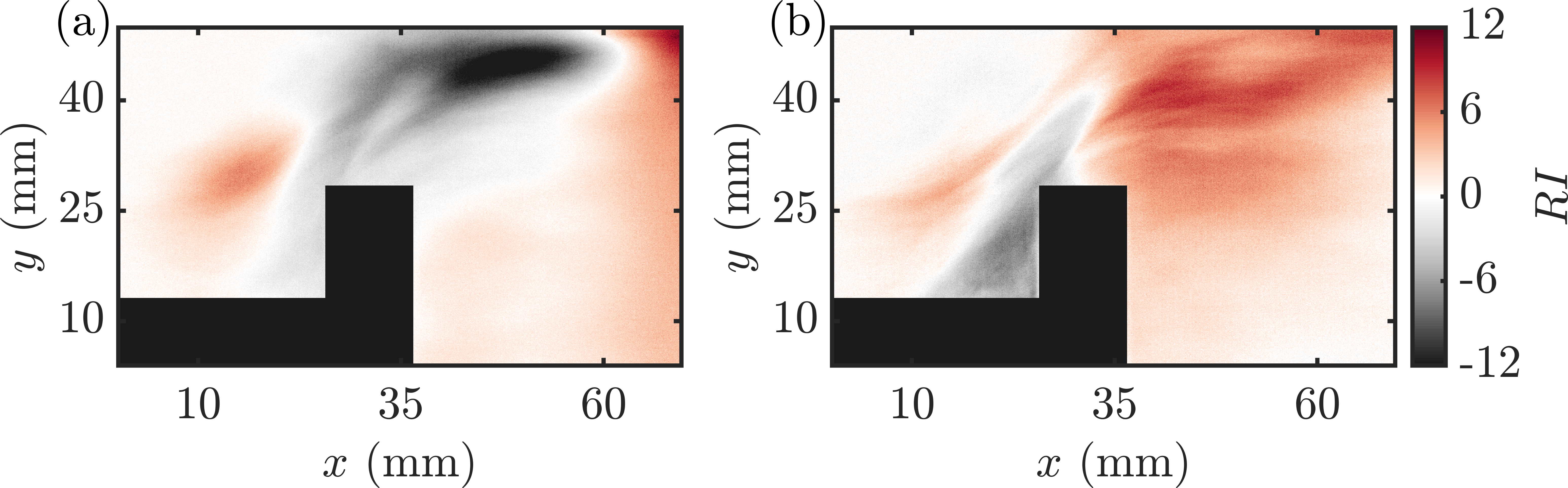}
    \caption{Spatial variation of Rayleigh index $RI$ plotted during epochs of (a) high-amplitude periodic and (b) low-amplitude aperiodic fluctuations in $p'$ }
    \label{fig_RI_spatial}
\end{figure}

The Rayleigh index ($RI$) is calculated as given by equation \ref{eq_RI}, where the integral over $t\in T$ represents conditional integration performed over all the epochs of high (or low) amplitude $p'$ dynamics identified by the method described in Section 3.2. When $RI$ is calculated as a conditional averaged during the epochs of high-amplitude $p'$ dynamics, we denote it as $RI_h$, and similarly as $RI_l$ corresponding to low-amplitude $p'$ dynamics.

\begin{equation}\label{eq_RI}
    RI(x,y)=\frac{1}{|T|} \int_{t\in T} p'(t)\dot{q}'(x,y,t) dt
\end{equation}
Here, $|T|$ is the total length of the epochs of high (or low) amplitude $p'$ dynamics. In figure \ref{fig_RI_spatial}(a,b), we plot the spatial variation of the Rayleigh index $RI_h$ and $RI_l$ corresponding to high and low amplitude $p'$ dynamics, respectively.

We find that $RI_h$ is positive primarily in a spatial pocket in the outer shear layer region (OSL) upstream of the bluff-body and negative in parts of the inner shear layer (ISL) and wake region above the bluff-body (see figure \ref{fig_RI_spatial}(a)). Also, $RI_h$ is negative but close to zero in the secondary flow recirculation (SFR) region in figure \ref{fig_RI_spatial}(a). Further, during the epochs of low-amplitude fluctuations in $p'$, we find that the conditionally averaged Rayleigh index $RI_l$ is positive in the OSL, and in the wake region above and close to the bluff-body, but negative in the SFR and ISL regions upstream of the bluff-body (see figure \ref{fig_RI_spatial}(b)). Also, during low-amplitude fluctuations in $p'$, thermoacoustic power sources in OSL decrease.

Clearly, a coherent region of strong sources of thermoacoustic power (i.e., positive Rayleigh index) exist in the OSL during the epochs of high-amplitude oscillations in $p'$. The spatial extent of these sources in the OSL decreases during epochs of low amplitude $p'$ dynamics. We note that, this spatial pocket is associated with the part of the flame that is anchored along OSL.  Local mixing is enhanced in the OSL owing to fluid entrainment and strong recirculation created in the flow by the relatively larger vortices that are shed in the recirculation zone during the high-amplitude $p'$ dynamics. From our analysis presented in figure 6(c), we discover that the frequency of local $\dot{q}'$ dynamics increases in locations in the OSL where coherent thermoacoustic power sources occur during the epochs of high-amplitude $p'$ dynamics.

On comparing $RI_h$ and $RI_l$ in SFR and ISL region adjacent to the bluff-body, we find that sinks of thermoacoustic power exist during both high and low amplitude $p'$ dynamics. However, the strength and spatial coherence of thermoacoustic power sinks increases during the epochs of low-amplitude $p'$ dynamics. These regions exhibit positive FM (evident in figure 6(c)).

Further, we find a coherent pocket of thermoacoustic power sinks during high-amplitude $p'$ dynamics, and thermoacoustic power sources during low-amplitude $p'$ dynamics in a spatial pocket above and downstream of the bluff body. Close to the top wall of the combustor, the large vortices impinge and undergo further combustion and the flame front distortions are significant. Thus, we can infer that the heat release rate fluctuations in these regions are out of phase with $p'$ during high-amplitude oscillations but become in phase with $p'$ during the low-amplitude fluctuations in $p'$. These regions exhibit negative FM in local $\dot{q}'$ dynamics as evident in figure 6(c).  

In summary, we infer that regions where positive FM occurs, we find stronger sources or weaker sinks of thermoacoustic power during epochs of high-amplitude $p'$ dynamics as compared to epochs of low-amplitude $p'$ dynamics.


\section{Mean heat release rate dynamics during experiments with continuous variation in the airflow rate}\label{app_rate_meanHRR}

\begin{figure}
    \centering
    \includegraphics[width=1\linewidth]{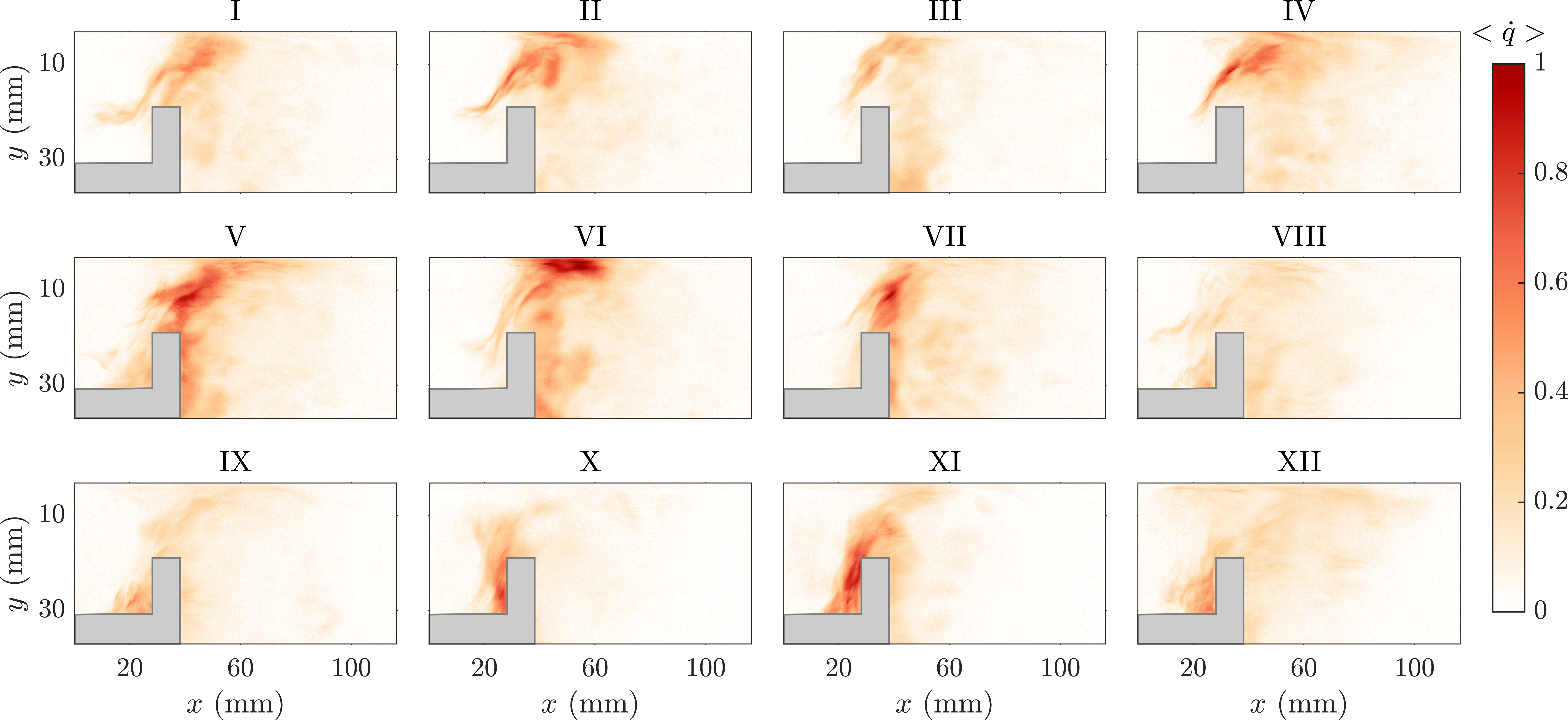}
    \caption{Spatial pattern mean heat release rate ($\langle\dot{q}\rangle$) averaged over short time frames of $0.6$ s, corresponding to the plots of $K_{lh}$ shown in figure \ref{fig_rate_results}. The time frames I-XII in this figure correspond to the time frames shown in figure \ref{fig_rate_results}. The pattern of mean heat release rate delineate the flame dynamics during each short time frame.}
    \label{fig_app_meanHRRrate}
\end{figure}

Figure \ref{fig_app_meanHRRrate} shows the spatial pattern of mean heat release rate ($\langle\dot{q}\rangle$) averaged over short time frames of $0.6$ s. The time frames indicated in figure \ref{fig_app_meanHRRrate} correspond to those in figure \ref{fig_rate_results}. The spatial pattern of average heat release rate indicates the flame dynamics during each epoch. We note that, heat released is maximum in regions where the reactants and the hot products mix efficiently. Usually these regions are associated with flame front oscillations as the vortices containing the unburnt reactants distort the flame where hot products are contained after combustion reaction. The flame anchoring and flame front oscillations can hence be conjectured using heat release rate dynamics.  

For example, during time frames (I-IV) in figure \ref{fig_app_meanHRRrate}, we see in experiments that the flame anchors predominantly close to the backward facing step and the flame front exists in the OSL and ISL. Thus, we find that on an average maximum heat release rate occurs in the OSL in figure \ref{fig_app_meanHRRrate}(I-IV). These are the time frames we classified as low-amplitude intermittency regime.

Next, during the moderate-amplitude intermittency regime, corresponding to the time frames (V-VIII) shown in figure \ref{fig_app_meanHRRrate}, we find that the flame anchors not only in the OSL close to the backward-facing step, but also in the ISL along the shaft upstream of the bluff-body. The flame front oscillates in the OSL, ISL and SFR regions, which are also the regions of maximum mean heat release rate in figure \ref{fig_app_meanHRRrate}(V and VIII). Also, note that the flame anchoring position can change along the shaft during some epochs. As a result, the flame can anchor in the SFR region and the flame front exists primarily in the ISL. Hence, we see that during certain time frames, the mean heat release rate is maximum only in the ISL and OSL and not in the SFR region such as in figure \ref{fig_app_meanHRRrate}(VI and VII). 

Finally, during time frames (IX-XII) in figure \ref{fig_app_meanHRRrate} corresponding to high-amplitude intermittency regime, we find that the maximum of mean heat release rate oscillations occurs predominantly in the SFR and ISL regions. As noted earlier, the flame anchors along the shaft of the bluff-body and not in the OSL and the flame anchors along the shaft at different locations during different epochs. In fact, the location of flame anchoring appears to oscillate to and fro along the shaft due to the periodic shedding of very large vortices in the dump plane. When the flame anchors upstream of the SFR region, then the SFR region is associated with flame front oscillations. In contrast, when the flame anchors in the SFR region, the flame front exists primarily in the OSL and wake of the bluff-body. 

As a result, we obtain high values of mean heat release rate in the SFR region as well as in the OSL, ISL and wake downstream of the bluff-body. The dynamic nature of flame anchoring can only be seen in experiments but cannot be seen in the mean heat release rate images. To characterize such dynamics, high-resolution Mie-scattering imaging is required which is a task for of future research.


\section{Classification of the three dynamical regimes during the occurrence of intermittency in experiments with continuous variation in the airflow rate}\label{app_classifyregimes}

\begin{figure}
    \centering
    \includegraphics[width=0.6\linewidth]{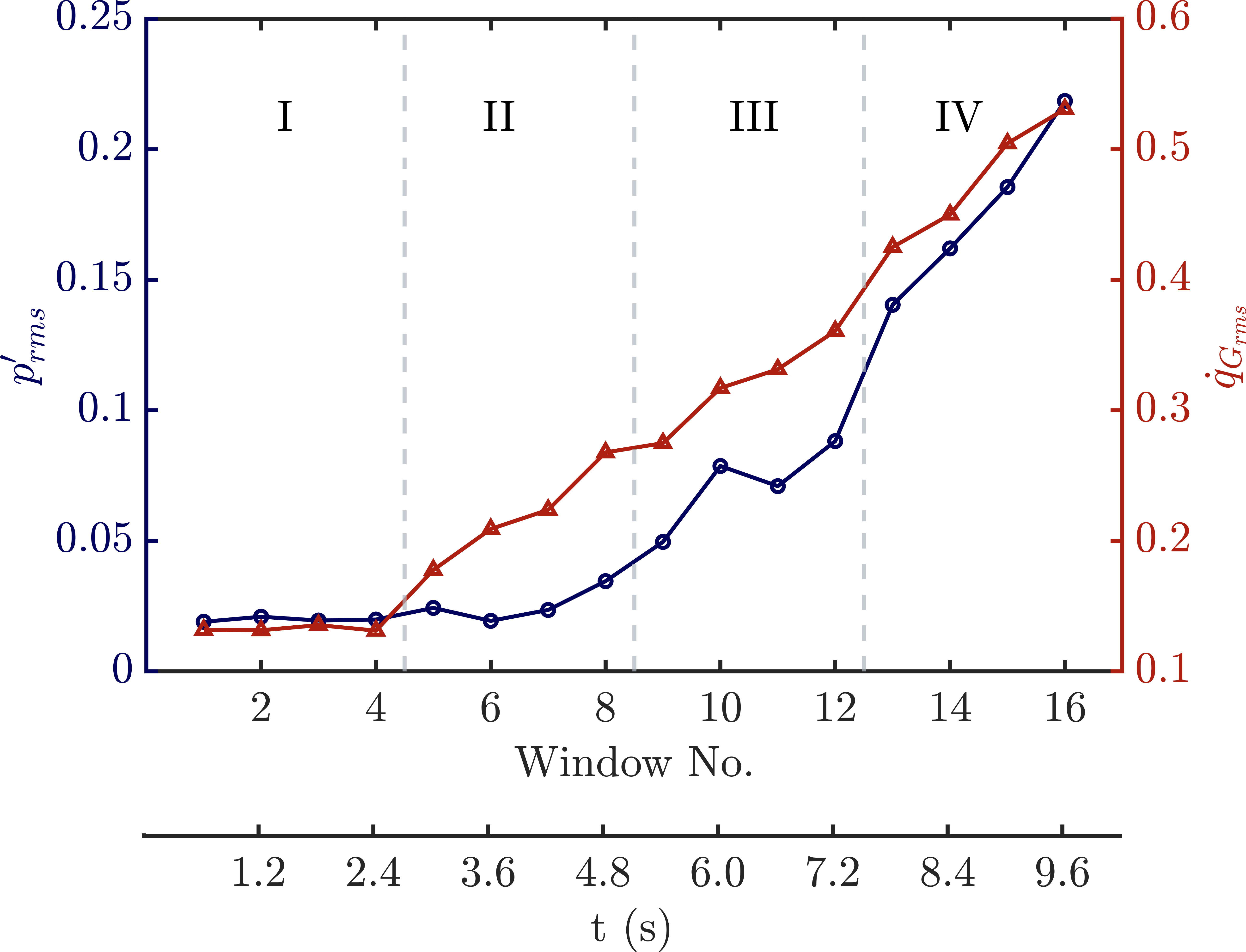}
    \caption{Variation of  $p'_{rms}$ and  $\dot{q}_{G_{rms}}$ obtained for a time-window of $T_w=0.6$ s during experiments in a turbulent bluff-body stabilized combustor, where the inlet airflow rate is varied continuously.}
    \label{fig_amp_demarc}
\end{figure}

In order to quantitatively distinguish the three intermittency regimes used in the main text, we consider variations in the root mean square of $p'$ and $\dot{q}_G$ fluctuations. In figure \ref{fig_amp_demarc} we plot the variation of $p'_{rms}$ and $\dot{q}_{G_{rms}}$ calculated for time-windows $T_w=0.6$ s. We perform this analysis for the dynamics in $t \in [0,9.6]$ s corresponding to $16$ time-windows as considered in the main text. 

From figure \ref{fig_amp_demarc}, we observe that during the epoch corresponding to the first four time-windows ($t\in[0,2.4]$ s, regime I in figure \ref{fig_amp_demarc}), the mean amplitude of $p'$ and $\dot{q}_G$ fluctuations is very low and the values of $p'_{rms}$ and $\dot{q}_{G_{rms}}$ are also very low. We classify these four time-windows as low-amplitude intermittency regime. Next, during $t\in[2.4,4.8]$ s, we find that the value of $\dot{q}_{G_{rms}}$ increases notably while the value of $p'_{rms}$ is still low and the increase is small. This dynamical regime corresponds to the time-windows $5-8$ (regime II in figure \ref{fig_amp_demarc}) and are characterized by relatively longer bursts of periodicity in $p'$ as compared to that in regime I. We classify this regime as the moderate-amplitude intermittency regime.

Finally, during $t\in[4.8,7.2]$ s (time-windows $9-12$, regime III in figure \ref{fig_amp_demarc}), we find that the values of both $p'_{rms}$ and  $\dot{q}_{G_{rms}}$ increase significantly and continuously. Note that, the fluctuations in $p'$ exhibit intermittent bursts of periodicity amidst relatively low-amplitude aperiodic fluctuations during this epoch. We thus classify this regime of  $t\in[4.8,7.2]$ s as the high-amplitude intermittency regime. For  $t \in [7.8,9.6]$ s (regime IV in figure \ref{fig_amp_demarc}),  $p'_{rms}$ and  $\dot{q}_{G_{rms}}$ continue to increase; however, $p'$ exhibits limit cycle oscillations.

\section{Effect of the length of time-window $T_w$ used for analysis of intermittency during experiments with continuous variation of the airflow rate }\label{app_Tw}

For experiments with continuous variation of the flow parameter, we construct the visibility graph form local heat release rate signals for short time-windows of $T_w$ length. In section \ref{sec_result_rate}, $T_w=0.6$ s. The epochs of high and low amplitude $p'$ fluctuations are distinguished via short-window FFT performed over $L=5$ cycles of the dominant acoustic mode. In order to find the effect of the parameter $T_w$, we select three epochs of different $T_w$ with maximum overlap during the three different dynamical regimes of intermittency obtained during experiments with continuous variation of the flow parameter. 

\begin{figure}
    \centering
    \includegraphics[width=0.9\linewidth]{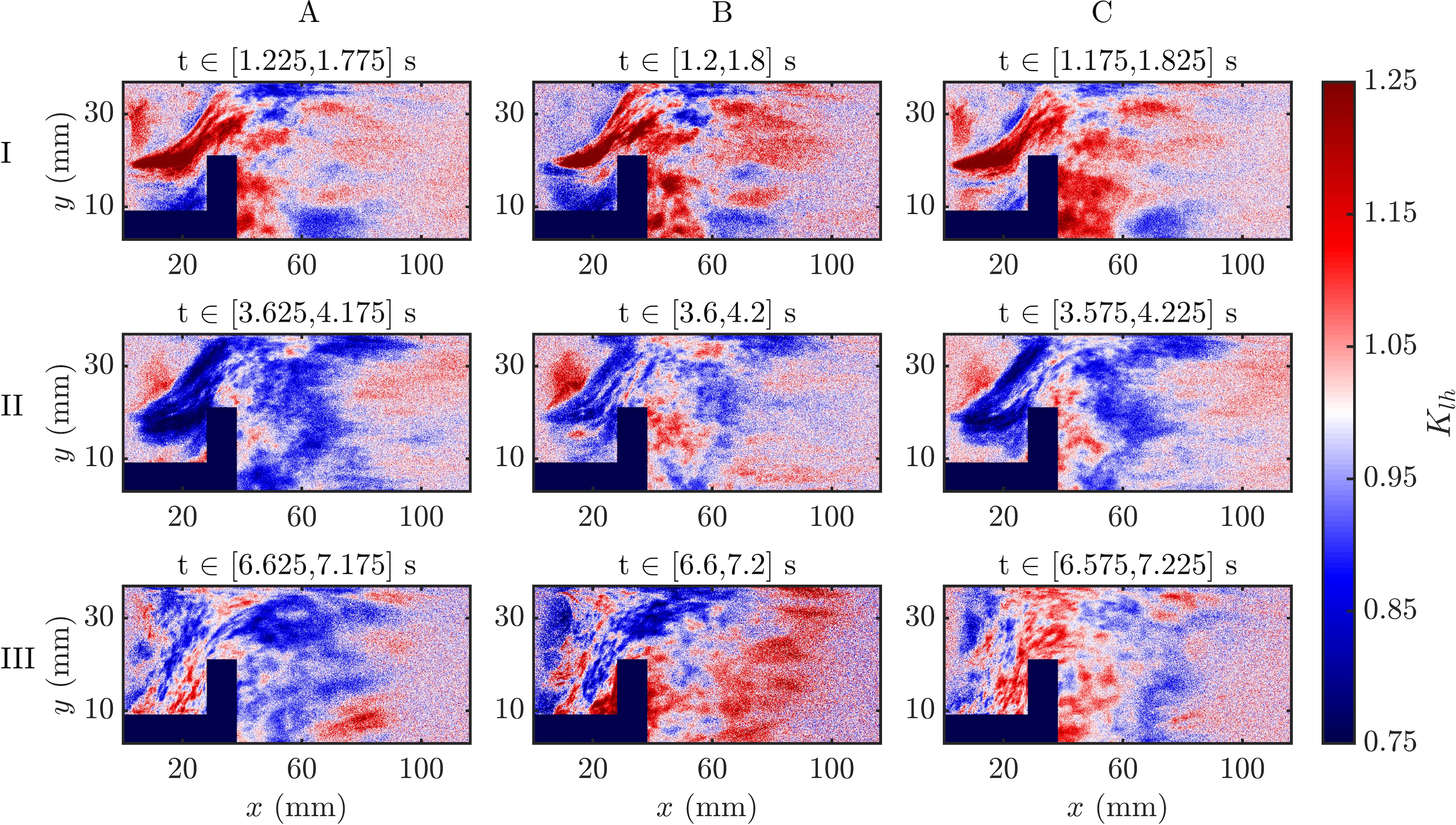}
    \caption{Spatial distribution of the ratio $K_{lh}$ obtained during the state of intermittency observed in experiments with continuous variation of the inlet airflow rate during low-amplitude (row I), moderate (row II) and high-amplitude (row III) intermittency dynamics for different values of $T_w=0.55,0.6,0.65$ s in columns A, B, C respectively. The epochs ($[t_1,t_2]$) selected are indicated in the title of each subplot. We observe that the spatial pockets of positive and negative FM identified during different regimes are similar for different $T_w$.}
    \label{fig_varyTw}
\end{figure}

For example, during low-amplitude intermittency dynamics (row-I in figure \ref{fig_varyTw}), we select three epochs and plot the spatial pattern of $K_{lh}$ during $t\in [1.225,1.775]$ s ($T_w=0.55$ s) in figure \ref{fig_varyTw}(I-B), $t\in [1.2,1.8]$ s ($T_w=0.6$ s) in figure \ref{fig_varyTw}(I-B), and $t\in [1.175,1.825]$ s ($T_w=0.65$ s) in figure \ref{fig_varyTw}(I-B). Similarly, epochs of different lengths ($T_w=0.55, 0.6, 0.65$ s) but with maximum overlap are selected during moderate and high-amplitude intermittency dynamics and the spatial plots of $K_{lh}$ are shown in row-II and row-III in figure \ref{fig_varyTw}, respectively.

Clearly, the spatial pockets of positive and negative FM identified during the epochs of different lengths ($T_w$) are similar during all the three dynamical regimes of intermittency. Our inference of the physical processes involved in the flame-acoustic interactions are therefore not very sensitive to $T_w$. We also note that, if we choose a very large time-window for analysis, the variations of the mean amplitude of $p'$ within the time-window will become significant. On the other hand, if a much smaller length of the time-window is used for the analysis, the statistics of the network measures may not be reliable.

\bibliographystyle{jfm}
\bibliography{ref}
\end{document}